\documentstyle[aaspp4]{article}

\received{}
\accepted{}
\journalid{VOL}{JOURNAL DATE}
\articleid{START PAGE}{END PAGE}
\paperid{MANUSCRIPT ID}

\begin{document}

%\magnification=1200
%\def\pmb#1{\setbox0=\hbox{$#1$}
% \kern-.025em\copy0\kern-\wd0
% \kern-.05em\copy0\kern-\wd0
% \kern-.025em\raise.0433em\box0 }
\def\gtsim {>\kern-1.2em\lower1.1ex\hbox{$\sim$}}
\def\ltsim {<\kern-1.2em\lower1.1ex\hbox{$\sim$}}
\def\ref{\hangindent=1pc \hangafter=1 \noindent}
\def\pmb#1{\mbox{\boldmath$#1$}}

\title{\bf R-modes of neutron stars with the superfluid core}

\author{Umin Lee$^1$ \& Shijun Yoshida$^{1,2}$}

\affil{$^1$Astronomical Institute, Tohoku University, Sendai, Miyagi 980-8578, Japan
\\ lee@astr.tohoku.ac.jp, yoshida@astr.tohoku.ac.jp}

\affil{$^2$Centro Multidisciplinar de Astrof\'{\i}sica -- CENTRA,
Departamento de F\'{\i}sica, Instituto Superior T\'ecnico, Av. Rovisco 
Pais 1,
1049-001 Lisboa, Portugal}

\begin{abstract}

We investigate the modal properties of the $r$-modes 
of rotating neutron stars with the core filled with neutron and proton superfluids,
taking account of entrainment effects between the superfluids.
The stability of the $r$-modes against gravitational radiation reaction
is also examined considering
viscous dissipation due to shear and a damping mechanism called mutual 
friction between the superfluids in the core.
We find the $r$-modes in the superfluid core
are split into ordinary $r$-modes and
superfluid $r$-modes, which we call, respectively, $r^o$- and $r^s$-modes.
The two superfluids in the core flow together for the $r^o$-modes, while
they counter-move for the $r^s$-modes.
For the $r^o$-modes, the coefficient $\kappa_0\equiv\lim_{\Omega\rightarrow 0}\omega/\Omega$ 
is equal to $2m/[l^\prime(l^\prime+1)]$, almost independent of the parameter $\eta$ that
parameterizes the entrainment effects between the superfluids,
where $\Omega$ is the angular frequency of rotation, 
$\omega$ the oscillation frequency observed in the corotating frame of the star, and
$l^\prime$ and $m$ are the indices of
the spherical harmonic function representing the angular dependence of the $r$-modes.
For the $r^s$-modes, on the other hand, $\kappa_0$ is equal to $2m/[l^\prime(l^\prime+1)]$
at $\eta=0$ (no entrainment), and it almost linearly increases as $\eta$ is increased from $\eta=0$.
The $r^o$-modes, for which $\pmb{w}^\prime\equiv\pmb{v}^\prime_p-\pmb{v}^\prime_n\propto\Omega^3$,
correspond to the $r$-modes discussed by Lindblom \& Mendell (2000), where
$\pmb{v}^\prime_n$ and $\pmb{v}^\prime_p$ are the Eulerian velocity perturbations
of the neutron and proton superfluids, respectively.
The mutual friction in the superfluid core is found ineffective to stabilize the $r$-mode
instability caused by the $r^o$-mode except in a few narrow regions of $\eta$.
The $r$-mode instability caused by the $r^s$-modes, on the other hand, is extremely
weak and easily damped by dissipative processes in the star.

\end{abstract}


\section{Introduction}

One of the roles expected for the $r$-mode instability to play
(Andersson 1998, Friedman \& Morsink 1998) 
is deceleration of the spin of newly born hot neutron stars by emitting
gravitational waves that carry away the angular momentum of the star
(e.g., Lindblom et al 1998).
We know, however, that
among older and colder neutron stars as found in LMXB systems
there are many rapidly rotating neutron stars like a millisecond pulsar
(see, e.g, Phinney \& Kulkarni 1994).
This fact suggests the possibility that
the $r$-mode instability does not always work well
to spin down the rapid rotation of the stars.
For the $r$-modes in cold neutron stars with a solid crust, for example,
Bildsten \& Ushomirsky (2000) suggested a damping mechanism operating
in the viscous boundary layer at the interface between the solid crust and the
fluid core,
to explain the clustering of spin frequencies around the value of 300Hz 
for accreting neutron stars in LMXB systems (van der Klis 2000; see also
Andersson, Kokkotas, \& Stergioulas 1999).
For the modal properties of the $r$-modes in neutron stars
with a solid crust, see Yoshida \& Lee (2001), who
showed that the $r$-modes in the core are largely affected by resonance with
the toroidal sound modes propagating in the solid crust.

As neutron stars cool down below $T\sim 10^9{\rm K}$,  
neutrons and protons in the core are believed to
be in superfluid states (e.g., Shapiro \& Tuekolwsky 1983).
In a rotating system of superfluids, it is well known that scattering between 
vortices in the superfluids and normal fluid particles produces
dissipation called mutual friction (e.g., Khalatnikov 1965, Tilley \& Tilley 1990).
Therefore, for people who are interested in the $r$-modes instability, 
it was a serious concern whether the $r$-mode instability could survive the dissipation
due to mutual
friction in the superfluid core of cold neutron stars, e.g., in LMXBs.
It was Lindblom \& Mendell (2000) who first examined the damping effects of 
mutual friction in the core
on the $r$-mode instability, and concluded that the mutual friction could not be
strong enough to damp out the instability in most of the parameter domains 
which we are interested in.
In their analysis of the $r$-modes in neutron stars,
Lindblom \& Mendell (2000)
employed a perturbative method in which the spin angular frequency $\Omega$ is regarded
as the infinitesimal parameter to expand the eigenfrequencies and eigenfunctions
of the modes, and they looked for the $r$-modes with the scalings given by
$\beta^\prime\equiv\mu_p^\prime-\mu_n^\prime+m_e\mu_e^\prime/m_p\propto \Omega^4$ 
and $\pmb{w}^\prime\equiv\pmb{v}_p^\prime-\pmb{v}_n^\prime\propto\Omega^3$, where
$\mu_p$, $\mu_n$, and $\mu_e$ are
the chemical potentials of the proton, neutron,
and electron in the core, and $\pmb{v}_p$ and $\pmb{v}_n$ are the velocities of the
proton and neutron superfluids, and the prime $(^\prime)$ indicates the
Euler perturbation of the quantity.

Recently, Andersson \& Comer (2001) discussed the dynamics of superfluid neutron star cores, 
and confirmed an earlier result by Lee (1995)
that there are no $g$-modes propagating in the superfluid core.
They also applied their argument to the $r$-modes in the core 
filled with neutron and proton superfluids, and
suggested the existence of two distinct families of the $r$-modes in the core, i.e.,
$r$-modes for which
the neutrons and the protons flow together, and those
for which the neutrons and the protons are counter moving.
Lindblom \& Mendell (2000) considered the former family of the $r$-modes,
which are less strongly affected by the mutual friction than the latter.

In this paper, we employ a different method of calculation
to investigate the $r$-modes in rotating neutron stars, although
the basic equations describing the dynamics of superfluids in the core
are essentially the same as those given in Lindblom \& Mendell (1994).
Our method of solution is a variant of that used in Lee \& Saio (1986).
Because the separation of variables is not possible for perturbations in
rotating stars,
we expand the perturbations in terms of
spherical harmonic functions $Y_l^m(\theta,\phi)$ with different $l$'s for a given $m$.
We substitute the expansions into linearized basic equations to obtain a set of
simultaneous linear ordinary differential equations of the expansion coefficients, which is
to be solved as an eigenvalue problem of the oscillation frequency.
In this method, we do not have to assume apriori a form of solutions for the $r$-modes
in the lowest order of $\Omega$.
In \S 2 we present the basic equations employed in this paper
for the dynamics of superfluids in the core, and in \S 3
dissipation processes considered in this paper are described.
\S 4 gives numerical results, and \S 5 and \S 6 are for discussions and conclusions.

\section{Oscillation Equations in the Superfluid Core of Rotating Neutron Stars}

Microscopically, the superfluids in a rotating system
move irrotationally everywhere except within the core
of vortex lines.
Averaging over many vortices in the fluids, we may define the average superfluid velocities
$<\pmb{v}>$,
which can satisfy the usual relation for uniform rotation 
$\nabla\times <\pmb{v}>=2\pmb{\Omega}$ in the equilibrium state
(see, e.g., Feynman 1972).
In the following we simply use $\pmb{v}$, instead of $<\pmb{v}>$, to signify the 
superfluid velocities.
Hydrodynamic equations for a rotating superfluid based on the two-fluid model
with the normal fluid and superfluid components
are derived, for example, in Khalatnikov (1965).

We derive basic equations governing superfluid motions
in the neutron star core in the Newtonian dynamics, assuming uniform rotation of the star.
The core is assumed to be filled with neutron and proton superfluids and 
a normal fluid of electron.
We also assume perfect charge neutrality between the protons and electrons because
the plasma frequency is much higher than the oscillation frequencies considered in
this paper (see, e.g., Mendell 1991a).
Since the transition temperatures $T_c\sim 10^9$K to neutron and proton superfluids 
are much higher than the interior temperatures of old neutron stars (see, e.g., 
Epstein 1988), 
we assume that all the neutrons and protons in the core are in superfluid states and
the normal fluid components of the fluids can be ignored.

The basic hydrodynamic equations employed in this paper for the
neutron and proton superfluids in rotating neutron stars 
are essentially the same as those given in Mendell (1991a) and
Lindblom \& Mendell (1994).
In a fluid system in which two superfluids coexist, the entrainment between 
the two superfluid motions occurs because a Cooper pair of one
fluid particles is affected by the force field produced by the other fluid particles
(Andreev \& Bashkin 1975 for a system of $^3$He and $^4$He superfluids).
In the system of the neutron and proton superfluids in the core
the entrainment effects between them are mediated
by quantum mechanical nuclear force
between the neutrons and the protons (e.g., Alpar et al 1984).
Here, we introduce the entrainment effects in
mass conservation equations, which are given by
\begin{equation}
{\partial\rho_n\over\partial t}+\nabla\cdot\pmb{j}_n=0, 
\end{equation}
and
\begin{equation}
{\partial\rho_p\over\partial t}+\nabla\cdot\pmb{j}_p=0, 
\end{equation}
where $\rho_n$ and $\rho_p$ are the mass densities of the neutron and proton superfluids,
and the mass current vectors $\pmb{j}_n$ and $\pmb{j}_p$ are defined as
\begin{equation}
\pmb{j}_n=\rho_{nn}\pmb{v}_n+\rho_{np}\pmb{v}_p, 
\end{equation}
and
\begin{equation}
\pmb{j}_p=\rho_{pp}\pmb{v}_p+\rho_{pn}\pmb{v}_n, 
\end{equation}
where $\pmb{v}_n$ and $\pmb{v}_p$ denote the velocities of the neutron and the proton superfluids,
respectively, and the coefficients $\rho_{nn}$, $\rho_{np}$, $\rho_{pp}$, and $\rho_{pn}$
are defined to satisfy $\rho_n=\rho_{nn}+\rho_{np}$ and
$\rho_p=\rho_{pp}+\rho_{pn}$ and $\rho_{np}=\rho_{pn}$ under Galilean transformations.
The mass conservation equation for the electron fluid is given by
\begin{equation}
{\partial\rho_e\over\partial t}+\nabla\cdot\left(\rho_e\pmb{v}_e\right)=0,
\end{equation}
where $\rho_e$ and $\pmb{v}_e$ denote, respectively,
the mass density and the velocity of the electron fluid.
The velocity equation of the neutron superfluid is in an inertial frame given by
\begin{equation}
{\partial \pmb{v}_n\over\partial t}+\pmb{v}_n\cdot\nabla\pmb{v}_n=-\nabla(\mu_n+\Psi)
+{\rho_{np}\over\rho_n}\left(\pmb{v}_p-\pmb{v}_n\right)\times\left(\nabla\times\pmb{v}_n\right),
\end{equation}
where $\mu_n$ is the chemical potential of neutron per unit mass, and $\Psi$ is the 
gravitational potential.
The term proportional to $\rho_{np}$ on the right hand side of equation (6) represents
a drag force between neutrons and protons.
If we assume perfect charge neutrality of 
the proton and electron plasma, we may have
\begin{equation}
\pmb{j}_p/\rho_p=\pmb{v}_e, \quad {\rm and} \quad \rho_p/m_p=\rho_e/m_e,
\end{equation}
where $m_p$ and $m_e$ are the proton and the electron masses.
The velocity equation for the proton-electron fluid is then given by
\begin{equation}
{\partial\over\partial t}\left(\pmb{v}_p+{m_e\over m_p}\pmb{v}_e\right)
+\pmb{v}_p\cdot\nabla\pmb{v}_p
+{m_e\over m_p}{\pmb{v}_e}\cdot\nabla{\pmb{v}_e}
=-\nabla\left(\mu_p+{m_e\over m_p}\mu_e+\zeta\Psi\right)
-{\rho_{np}\over\rho_p}\left(\pmb{v}_p-\pmb{v}_n\right)\times\left(\nabla\times\pmb{v}_p\right), 
\end{equation}
where $\zeta=1+m_e/m_p$, and $\mu_p$ and $\mu_e$ are the chemical potentials
per unit mass for the proton and electron, respectively.
Note that we have neglected the entropy carried by the normal fluid of electron
for simplicity.
The Poisson equation is given by
\begin{equation}
\nabla^2\Psi=4\pi G\rho, 
\end{equation}
where $\rho=\rho_n+\rho_p+\rho_e$ and $G$ is the gravitational constant.

To linearise the hydrodynamic equations for the superfluids in
rotating neutron stars, we assume that
the neutron and proton superfluids and the electron normal fluid in an equilibrium state 
are in the same rotational motion
with the angular velocity $\Omega$ around the axis of rotation, which is along the z-axis.
In a perturbed state, however, the neutron and the proton superfluids move differently
from each other in the core, obeying their own governing equations.
The mass current vectors are linearized to be
\begin{equation}
\pmb{j}_n^\prime=\rho_n^\prime\pmb{v}_0+\tilde{\pmb{j}}_n^\prime,
\quad {\rm and} \quad
\pmb{j}_p^\prime=\rho_p^\prime\pmb{v}_0+\tilde{\pmb{j}}_p^\prime,
\end{equation}
where the prime $(^\prime)$ indicates the Eulerian perturbation of the quantity,
and $\pmb{v}_{n0}=\pmb{v}_{p0}=\pmb{v}_0=r\sin\theta\Omega\pmb{e}_\phi$
is the fluid velocity in the equilibrium state, and
\begin{equation}
\tilde{\pmb{j}}_n^\prime=\rho_{nn}\pmb{v}_n^\prime+\rho_{np}\pmb{v}_p^\prime,
\quad {\rm and} \quad
\tilde{\pmb{j}}_p^\prime=\rho_{pn}\pmb{v}_n^\prime+\rho_{pp}\pmb{v}_p^\prime
\end{equation}
are the perturbed mass current vectors in a corotating frame.
The perturbed superfluid velocities are then given in terms of $\tilde{\pmb{j}}_n^\prime$
and $\tilde{\pmb{j}}_p^\prime$ as
\begin{equation}
\pmb{v}_n^\prime={\rho_{11}\over\rho_n^2}\tilde{\pmb{j}}_n^\prime
+{\rho_{12}\over\rho_n\rho_p}\tilde{\pmb{j}}_p^\prime,
\quad {\rm and} \quad
\pmb{v}_p^\prime={\rho_{21}\over\rho_n\rho_p}\tilde{\pmb{j}}_n^\prime
+{\rho_{22}\over\rho_p^2}\tilde{\pmb{j}}_p^\prime,
\end{equation}
where
\begin{equation}
\rho_{11}={\rho_{pp}\rho_n^2\over\tilde{\rho}^2}, \quad
\rho_{22}={\rho_{nn}\rho_p^2\over\tilde\rho^2}, \quad 
\rho_{12}=\rho_{21}=-{\rho_{np}\rho_n\rho_p\over\tilde\rho^2},
\end{equation}
and
\begin{equation}
\tilde\rho^2=\rho_{nn}\rho_{pp}-\rho_{np}\rho_{pn}.
\end{equation}
Note that $\rho_{11}+\rho_{12}=\rho_n$ and $\rho_{22}+\rho_{21}=\rho_p$.

If the equilibrium structure is axisymmetric about the rotation axis, 
the time dependence and $\phi-$dependence of the perturbations can be given by
$\exp(i\sigma t+i m\phi)$ with $\sigma$ being the oscillation frequency observed 
in an inertial frame, and $m$ is an integer representing the azimuthal wavenumber.
Introducing the vectors $\pmb{\xi}^n$, $\pmb{\xi}^p$, and $\pmb{\xi}^e$
defined as
\begin{equation}
\pmb{\xi}^n\equiv{\tilde{\pmb{j}}_n^\prime\over i\omega\rho_n},
\quad
\pmb{\xi}^p\equiv{\tilde{\pmb{j}}_p^\prime\over i\omega\rho_p},
\quad {\rm and} \quad
\pmb{\xi}^e\equiv {\pmb{v}_e^\prime\over i\omega},
\end{equation}
where $\omega\equiv\sigma+m\Omega$ 
is the oscillation frequency observed in a
corotating frame of the star,
the mass conservation equations are linearised to be
\begin{equation}
\rho_n^\prime+\nabla\cdot\left(\rho_n\pmb{\xi}^n\right)=0,
\end{equation}
and
\begin{equation}
\rho_p^\prime+\nabla\cdot\left(\rho_p\pmb{\xi}^p\right)=0.
\end{equation}
Note that the mass conservation equation for the electron fluid becomes the same as that for
the proton fluid because of the assumption of perfect charge neutrality, that is, 
$\pmb{\xi}^p=\pmb{\xi}^e$, and $\rho_p^\prime/m_p=\rho_e^\prime/m_e$.
The velocity equations (6) and (8) are linearized as
\begin{equation}
{\rho_{11}\over\rho_n}\pmb{F}\left(\pmb{\xi}^n\right)+
{\rho_{12}\over\rho_n}\pmb{F}\left(\pmb{\xi}^p\right)=
-\nabla\left(\mu_n^\prime+\Psi^\prime\right)
+i\omega{\rho_{np}\rho_p\over\tilde\rho^2}\left(\pmb{\xi}^p-\pmb{\xi}^n\right)
\times\left(\nabla\times\pmb{v}_0\right),
\end{equation}
and
\begin{equation}
{\rho_{21}\over\rho_p}\pmb{F}\left(\pmb{\xi}^n\right)+
\left({\rho_{22}\over\rho_p}+{m_e\over m_p}\right)\pmb{F}\left(\pmb{\xi}^p\right)=
-\nabla\left(\tilde\mu^\prime+\zeta\Psi^\prime\right)
-i\omega{\rho_{np}\rho_n\over\tilde\rho^2}\left(\pmb{\xi}^p-\pmb{\xi}^n\right)
\times\left(\nabla\times\pmb{v}_0\right),
\end{equation}
where $\tilde\mu\equiv\mu_p+\mu_e{m_e}/m_p$, and 
\begin{equation}
\pmb{F}\left(\pmb{\xi}\right)\equiv-\omega\sigma\pmb{\xi}+i\omega\pmb{v}_0\cdot\nabla\pmb{\xi}
+i\omega\pmb{\xi}\cdot\nabla\pmb{v}_0.
\end{equation}
The Poisson equation is reduced to
\begin{equation}
\nabla^2\Psi^\prime=4\pi G\left(\rho_n^\prime+\zeta\rho_p^\prime\right).
\end{equation}
Using a variant of Gibbs-Duhem relation,
the pressure perturbation is given by
\begin{equation}
p^\prime=\rho_n\mu_n^\prime+\rho_p\tilde\mu^\prime, 
\end{equation}
where we have again ignored the entropy carried by the electron normal fluid.
Note also that the superfluids carry no entropy.

To obtain a relation between the densities $\rho_n$ and $\rho_p$ and the chemical potentials
$\mu_n$ and $\tilde\mu$, we begin with writing the energy density $e$ as
\begin{equation}
e=e\left(\rho_n,\rho_p\right), 
\end{equation}
with which the chemical potentials are defined as
\begin{equation}
\mu_n\left(\rho_n,\rho_p\right)=\left(\partial e/\partial \rho_n\right)_{\rho_p}, \quad
\mu_p\left(\rho_n,\rho_p\right)=\left(\partial e/\partial \rho_p\right)_{\rho_n}.
\end{equation}
If the chemical potential of the electron is given by
\begin{equation}
\mu_e(\rho_e)=c^2\sqrt{1+(3\pi^2\hbar^3\rho_e/m_e^4c^3)^{2/3}}, 
\end{equation}
we have, assuming $\rho_e=\rho_pm_e/m_p$,
\begin{equation}
\left(\matrix{\mu_n^\prime \cr \tilde\mu^\prime\cr}\right)
=\left(\matrix{{\cal P}_{11} & {\cal P}_{12}\cr {\cal P}_{21} & {\cal P}_{22}\cr}\right)
\left(\matrix{\rho_n^\prime \cr \rho_p^\prime \cr}\right),
\end{equation}
where
\begin{equation}
{\cal P}_{11}=\left({\partial\mu_n\over \partial\rho_n}\right)_{\rho_p}, \quad
{\cal P}_{12}=\left({\partial\mu_n\over\partial\rho_p}\right)_{\rho_n}
=\left({\partial\tilde\mu\over\partial\rho_n}\right)_{\rho_p}={\cal P}_{21}, \quad
{\cal P}_{22}=\left({\partial\tilde\mu\over\partial\rho_p}\right)_{\rho_n}.
\end{equation}
We write the inverse of equation (26) as
\begin{equation}
\left(\matrix{\rho_n^\prime \cr \rho_p^\prime \cr}\right)=
\left(\matrix{{\cal Q}_{11} & {\cal Q}_{12}\cr {\cal Q}_{21} & {\cal Q}_{22}\cr}\right)
\left(\matrix{\mu_n^\prime \cr \tilde\mu^\prime\cr}\right),
\end{equation}
where
\begin{equation}
\left(\matrix{{\cal Q}_{11} & {\cal Q}_{12}\cr {\cal Q}_{21} & {\cal Q}_{22}\cr}\right)
=\left(\matrix{{\cal P}_{11} & {\cal P}_{12}\cr {\cal P}_{21} & {\cal P}_{22}\cr}\right)^{-1}.
\end{equation}

To represent the rotationally deformed equipotential surfaces of a rotating star,
we employ a coordinate system $(a,\theta,\phi)$, the relation of which
to spherical polar coordinates $(r,\theta,\phi)$ is given by
$r=a[1+\epsilon(a,\theta)]$,
where $\epsilon$ is proportional to $\Omega^2$ and represents a small deviation of the
equipotential surface from the corresponding spherical equipotential surface of 
the non-rotating star.
We apply Chandrasekhar-Milne expansion (Chandrasekhar 1933a,b, 
Tassoul 1978) to the hydrostatic equations to determine the function $\epsilon$
in the form $\epsilon(a,\theta)=\alpha(a)+\beta(a)P_2(\cos\theta)$ 
with $P_2$ being the Legendre function.
See Lee (1993) for the definition of $\alpha(a)$ and $\beta(a)$.
Since for uniformly rotating stars
all the physical quantities in hydrostatic equilibrium 
are constant on the deformed equipotential surface,
labeled by the coordinate $a$, we write
the linearised basic equations using the coordinates $(a,\theta,\phi)$
(Saio 1981, Lee 1993).
In our formulation, the terms up to of order of $\Omega^3$ in the perturbed velocity 
equations are retained so that the eigenfrequencies of the $r$-modes are correctly
determined to the order of $\Omega^3$ (see Yoshida \& Lee 2000a).

The perturbations in a uniformly rotating star are expanded in terms of spherical harmonic
functions with different $l$'s for a given $m$ (e.g., Lee \& Saio 1986).
For example, the vector $\pmb{\xi}^n$ is given by
\begin{equation}
{\xi_a^n\over a}=\sum_{l\geq |m|} S_l^n(a)Y_l^m(\theta,\phi)e^{i\sigma t},
\end{equation}
\begin{equation}
{\xi_\theta^n\over a}=\sum_{l\geq|m|}\left[H_l^n(a){\partial\over\partial\theta}
Y_l^m(\theta,\phi)+T_{l^\prime}^n(a){1\over\sin\theta}{\partial\over\partial\phi}
Y_{l^\prime}^m(\theta,\phi)\right]e^{i\sigma t},
\end{equation}
\begin{equation}
{\xi_\phi^n\over a}=\sum_{l\geq|m|}\left[H_l^n(a){1\over\sin\theta}{\partial\over
\partial\phi}Y_l^m(\theta,\phi)-T_{l^\prime}^n(a){\partial\over\partial\theta}
Y_{l^\prime}^m(\theta,\phi)\right]e^{i\sigma t},
\end{equation}
and the Euler perturbation of the gravitational potential, $\Psi^\prime$ is given as
\begin{equation}
\Psi^\prime=\sum_{l\geq|m|}\Psi^\prime_l(a)Y_l^m(\theta,\phi)
e^{i\sigma t},
\end{equation}
where $l=|m|+2(k-1)$ and $l'=l+1$ for even modes, and $l=|m|+2k-1$ and $l'=l-1$
for odd modes, and $k=1,~2,~3,~\cdots$.
Substituting these expansions and the like
into the linearised basic equations (16) $\sim$ (19) and (21), 
we obtain oscillation equations given as a set of
simultaneous linear ordinary differential equations of the expansion coefficients
(see Appendix), which is to be integrated in the superfluid core.
The oscillation equations solved in the normal fluid envelope are the same as those
given in Yoshida \& Lee (2000a).

To obtain a complete solution of an oscillation mode, solutions in
the superfluid core and in the normal fluid envelope are matched at the interface between 
the two domains by imposing jump conditions given by
\begin{equation}
\xi_a=\xi_a^n, \quad \xi_a=\xi_a^p, \quad
[p^\prime]^+_-=0, \quad [\Psi^\prime]^+_-=0,
\quad {\rm and} \quad [d\Psi^\prime/da]^+_-=0, 
\end{equation}
where 
$[f(x)]^+_-\equiv\lim_{s\rightarrow 0}\{f(x+s)-f(x-s)\}$.
The boundary conditions at the stellar center is 
the regularity condition of the perturbations
$\xi_a^n$, $\xi_a^p$, $\mu_n^\prime/g$, $\tilde\mu^\prime /g$, and $\Psi^\prime/g$, 
where $g=GM_a/a^2$.
The boundary conditions at the stellar surface 
are $\delta p=0$ and $\Psi_l^\prime\propto a^{-(l+1)}$,
where $\delta p$ is the Lagrangian perturbation of the pressure.

For numerical computation, oscillation equations of a finite dimension are obtained
by disregarding the terms
with $l$ larger than $l_{max}=|m|+2k_{max}-1$ in the expansions of perturbations
such as given by (30) to (33).
For the $r$-modes with $l^\prime=|m|$ calculated in this paper,
we usually use $k_{max}=6$ so that we can get reasonable convergence of
the eigenfrequency and the eigenfunction.
We solve the oscillation equations of a finite dimension as an eigenvalue problem of the 
oscillation frequency $\omega$ using a Henyey type relaxation method
(see, e.g., Unno et al 1989).

\section{Dissipations}

The stability of an oscillation mode of a star is determined by summing up all contributions 
from various damping and excitation mechanisms.
If we consider the contributions from gravitational radiation reaction, 
viscous processes, and mutual friction in the superfluid core,
the energy loss (or gain) rate $dE/dt$ of a normal mode in a rotating neutron
star may be given by 
\begin{eqnarray}
{dE\over dt}=&&-\sigma\omega\sum_{l=2}^\infty N_l\sigma^{2l}
\left(\left|D_{lm}^\prime\right|^2+\left|J_{lm}^\prime\right|^2\right)
-\int d^3x \left(\sum_{ij}{\sigma^{\prime ij}
\sigma_{ij}^{\prime*}\over 2\zeta_S}+\zeta_B\left|\nabla\cdot\pmb{v}^\prime\right|^2\right) 
\nonumber \\
&&-2\Omega\int d^3x\rho_nB_n\left({\tilde\rho^2\over\rho_n\rho_p}\right)^2
\left(\pmb{w}^\prime\cdot\pmb{w}^{\prime *}-w_z^\prime w_z^{\prime *}\right) \nonumber\\
=&& \left({dE\over dt}\right)_{GD}+\left({dE\over dt}\right)_{GJ}+\left({dE\over dt}\right)_{S}
+\left({dE\over dt}\right)_{B}+\left({dE\over dt}\right)_{MF},
\end{eqnarray}
where the asterisk $(^*)$ indicates the complex conjugate of the quantity, and
the canonical energy $E$ of oscillation observed in the corotating frame of the star is
defined in the normal fluid envelope as (Friedman \& Schutz 1978)
\begin{equation}
E={1\over 2}\int d^3x \left(\rho\pmb{v}^\prime\cdot\pmb{v}^{\prime *}+
{p^\prime\over\rho}\rho^{\prime *}-{\nabla\Psi^\prime\cdot\nabla\Psi^{\prime *}\over 4\pi G}
\right),
\end{equation}
and in the superfluid core as (Mendell 1991b)
\begin{equation}
E={1\over 2}\int d^3x\left(\rho\pmb{v}^\prime\cdot\pmb{v}^{\prime *}+
{\tilde\rho^2\over\rho}\pmb{w}^\prime\cdot\pmb{w}^{\prime *}+
\sum_{ij}{\cal P}_{ij}\rho^\prime_i\rho^{\prime *}_j
-{\nabla\Psi^\prime\cdot\nabla\Psi^{\prime *}\over 4\pi G}\right),
\end{equation}
where 
$\pmb{v}^\prime=(\rho_n\pmb{v}_n^\prime+\rho_p\pmb{v}_p^\prime)/\rho$, 
$\pmb{w}^\prime=\pmb{v}_p^\prime-\pmb{v}_n^\prime$,   
$\rho^\prime_1=\rho^\prime_n$, and $\rho^\prime_2=\rho^\prime_p$ in the core.

The terms $(dE/dt)_{GD}$ and $(dE/dt)_{GJ}$ on the right hand side of equation (35)
denote the energy loss (or gain) rates due to gravitational
radiation reaction associated with the mass multipole moment $D_{lm}^\prime$ and
the mass current multipole moment $J_{lm}^\prime$, where
\begin{equation}
D_{lm}^\prime=\int d^3x\rho^\prime r^l Y_l^{m*}, 
\end{equation}
\begin{equation}
J_{lm}^\prime={2\over c(l+1)}
\int d^3x r^l\left(\rho\pmb{v}^\prime+\rho^\prime\pmb{v}_0\right)\cdot
\left(\pmb{r}\times\nabla Y_l^{m*}\right),
\end{equation}
and
\begin{equation}
N_l={4\pi G\over c^{2l+1}}{(l+1)(l+2)\over l(l-1)[(2l+1)!!]^2}, 
\end{equation}
and $c$ is the velocity of light (Thorne 1980, Lindblom et al 1998).

The terms $(dE/dt)_{S}$ and $(dE/dt)_B$ are the energy dissipation rates 
due to shear and bulk viscosities, 
and $\zeta_S$ and $\zeta_B$ are the shear and the bulk viscosity coefficients, 
and $\sigma_{ij}^\prime$ is 
the traceless stress tensor for the perturbed velocity field
(e.g., Landau \& Lifshitz 1987).
In this paper, we ignore the contribution from the bulk viscosity, which is
important only for newly born hot neutron stars without superfluids in the core.
The shear viscosity coefficient we use in the superfluid core is
\begin{equation}
\zeta_S=6\times 10^{18}\left({\rho\over 10^{15}{\rm g/cm^3}}\right)^2
\left({10^9 K\over T}\right)^2 {\rm g/cm ~s}
\end{equation}
(Cutler \& Lindblom 1987, Sawyer 1989), and that in the normal fluid envelope is given by
\begin{equation}
\zeta_S=2\times 10^{18}\left({\rho\over 10^{15}{\rm g/cm^3}}\right)^{9/4}
\left({10^9 K\over T}\right)^2 {\rm g/cm ~s}
\end{equation}
(Cutler \& Lindblom 1987, Flowers \& Itoh 1979).
The stress tensor $\sigma_{ij}^\prime$ is evaluated by using $\pmb{v}_e^\prime$ 
in the superfluid core (Lindblom \& Mendell 2000).

The term $(dE/dt)_{MF}$ is the energy loss rate due to 
mutual friction in the superfluid core, 
and the dimensionless coefficient $B_n$ is given by (Mendell 1991b)
\begin{equation}
B_n=0.011\times {\rho_p\over\rho_n}
\left({\rho_{pp}\over\rho_{p}}\right)^{1/2}
\left({\rho_{pn}\over\rho_{pp}}\right)^2
\left({\rho_p\over 10^{14}{\rm gcm^{-3}}}\right)^{1/6}.
\end{equation}
Mutual friction is a dissipation mechanism inherent to a rotating system 
of superfluids, and it is caused by scattering of normal fluid particles off the vortices
in the superfluids.
Since we have assumed perfect charge neutrality between the electrons and protons, we consider
scattering between the normal electrons and vortices of the neutron superfluid
(Mendell 1991b, see also Alpar etal 1984).

As is indicated by the first two terms on the right hand side of equation (35), 
if a normal mode has an oscillation frequency that satisfies
$-\sigma\omega>0$, 
the oscillation energy $E$ in the corotating frame, in the absence of other damping mechanisms, 
increases as a result of
gravitational wave radiation, indicating instability of the mode (Friedman \& Schutz 1978).
It was Andersson (1998) and Friedman \& Morsink (1998) who realized that the $r$-modes have
oscillation frequencies that satisfy the instability condition.

The damping (or growth) time-scale $\tau$ of a normal mode may be given by
\begin{equation}
{1\over\tau}={1\over 2E}\left({dE\over dt}\right)={1\over\tau_{GD}}+{1\over\tau_{GJ}}+
{1\over\tau_S}+{1\over\tau_{MF}},
\end{equation}
where $\tau_i=2E/(dE/dt)_i$. For the $r$-modes of $l^\prime =|m|$, it is convenient to
derive an extrapolation formula of the time-scale $\tau$ given as a function of $\Omega$
and the interior temperature $T$ (e.g., Lindblom et al 1998, Lindblom \& Mendell 2000):
\begin{equation}
{1\over\tau (\Omega,T)}
={1\over\tau^0_{GD}}\left({\Omega^2\over \pi G\bar\rho}\right)^{l+2}
+{1\over\tau^0_{GJ}}\left({\Omega^2\over \pi G\bar\rho}\right)^{l}
+{1\over\tau^0_S}\left({10^9 K\over T}\right)^2
+{1\over\tau^0_{MF}}\left({\Omega^2\over \pi G\bar\rho}\right)^\gamma,
\end{equation}
where $\bar\rho=M/(4\pi R^3/3)$, and 
only the dominant term in each of the dissipation processes with $l=|m|+1=3$
has been retained for the $r$-modes. 
The quantities $\tau^0_{GD}$, $\tau^0_{GJ}$, $\tau^0_S$, and $\tau^0_{MF}$
are assumed to be only weakly dependent on $\Omega$ and $T$.

\section{Numerical Results}

Following Lindblom \& Mendell (2000), 
we employ a polytropic model of index $N=1$ with the mass $M=1.4M_\odot$ and the radius $R=12.57$km
as a background model for modal analysis.
The model is divided into a superfluid core and a normal fluid
envelope, the interface of which is set at $\rho=\rho_s=2.8\times 10^{14}$g/cm$^3$.
In the normal fluid envelope, we use the polytropic equation of state given by $p=K\rho^2$
both for the equilibrium structure and for the oscillation equations.
The core is assumed to be filled with neutron and proton superfluids and a normal fluid of electron,
for which an equation of state, labeled
A18+$\delta v$+UIX (Akmal, Pandharipande, and Ravenhall 1998), is used to
give the energy density (23) and the relation (26) used for the oscillation equations.
For the mass density coefficients $\rho_{nn}$, $\rho_{pp}$, and $\rho_{np}$ in the core,
we employ an empirical relation given by (see Lindblom \& Mendell 2000)
\begin{equation}
{\rho_p/\rho}\approx 0.031+8.8\times 10^{-17}\rho, 
\end{equation}
and a formula given by
\begin{equation}
\rho_{np}=-\eta\rho_n, 
\end{equation}
where $\eta$ is a parameter of order of $\sim 0.04$ that parametrizes the entrainment effects
between the two superfluids
(Borumand, Joynt, \& Klu\'zniak 1996).

We find that the $r$-modes of $l^\prime=m$ in the superfluid core are split into
ordinary $r$-modes and superfluid $r$-modes, which we call $r^o$-modes and $r^s$-modes.
The toroidal components $iT_{l^\prime}$ of the $r^o$- and $r^s$-modes of $l^\prime=m=2$
are plotted versus $a/R$ for the case of $\eta=0.04$ and $\bar\Omega\equiv\Omega/\sqrt{GM/R^3}=0.01$ 
in Figure 1, where
the solid, dotted, dashed, and dash-dotted curves are used to indicate respectively
the toroidal components $iT_m^n$, $iT_{m+2}^n$, $iT_m^p$, and $iT_{m+2}^p$ 
in the superfluid core, and
the amplitude normalization is given by $\max(|iT_m|)=1$.
The figure shows that the two superfluids in the core flow together for the $r^o$-modes, while
they counter-move for the $r^s$-modes.
Note that the amplitudes of $iT_{m+2}^p$ for the $r^s$-mode
are not necessarily negligibly small compared with
those of $iT_m^p$ in the core.
This splitting of the $r$-modes in the superfluid core into two distinct families has been
suggested by Andersson \& Comer (2001).
For the $r^o$-mode, the radial dependence of the difference $iT_m^p-iT_m^n$ in the core
is given in Figure 2, which shows that the difference is quite small 
compared with the normalization $\max(|iT_m|)=1$.
It is important to note that
the $l^\prime=m$ $r$-modes having basically nodeless and dominant $iT_m$ are the 
only $r$-modes we can find, as in the case of
the $r$-modes in isentropic models (see Yoshida \& Lee 2000a,b).

In Table 1, the expansion coefficients $\kappa_0$ and $\kappa_2$, and the scaled
damping (or growth) timescales $\tau^0_i$ for the $l'=m=2$ $r$-modes are tabulated
for the cases of $\eta=0$, $0.02$, $0.04$, and $0.06$, 
where the coefficients $\kappa_0$ and $\kappa_2$ are defined in the expansion:
\begin{equation}
\omega/\Omega=\kappa_0(\eta)+\kappa_2(\eta)\bar\Omega^2+O(\bar\Omega^4).
\end{equation}
Note that the exponent $\gamma$ employed to define $\tau^0_{MF}$ in equation (45) is 
$\gamma=2.5$ for the $r^o$-modes 
and $\gamma=0.5$ for the $r^s$-modes for the non-zero $\eta$'s in the table.
This is because the velocity difference 
$\pmb{w}^\prime=\pmb{v}^\prime_p-\pmb{v}^\prime_n$ approximately
scales as $\pmb{w}^\prime\propto\Omega^3$ for
the $r^o$-modes and as $\pmb{w}^\prime\propto\Omega$ for the $r^s$-modes 
(see Lindblom \& Mendell 2000).
The coefficient $\kappa_0$ for the $r^o$-modes is numerically equal to 
$2m/[l^\prime(l^\prime+1)]$, 
and $\kappa_2$ is almost independent of the entrainment parameter $\eta$.
The value of $\kappa_0$ is the same as the value found by Lindblom \& Mendell (2000)
and the value of $\kappa_2$ differs only by 2.5\%, suggesting that the $r^o$-modes are
the same modes found by Lindblom \& Mendell (2000).
(Because of different normalization conventions the value of $\kappa_2$ reported by
Lindblom \& Mendell must be multiplied by a factor of $4/3$ before comparing
with our results.)
The coefficients $\kappa_0$ and $\kappa_2$ for the $r^s$-modes, on the other hand, 
appreciably depends on $\eta$, and $\kappa_0$ deviates from $2m/[l^\prime(l^\prime+1)]$ 
as $\eta$ is increased from $\eta=0$.
This kind of deviation of $\kappa_0$ from $2m/[l^\prime(l^\prime+1)]$ for the $l^\prime=m$ $r$-modes
has been found for relativistic neutron stars where
the relativistic factor $GM/c^2R$ is regarded as a parameter (Yoshida 2001; Yoshida \& Lee 2002a).
Computing the $r^s$-mode of $l^\prime=m=2$
as a function of $\eta$, we find that a linear formula given by
\begin{equation}
\kappa_0^s(\eta)\approx0.667+9.35\eta
\end{equation}
gives a good fit to the $r^s$-mode frequency except at avoided crossings with inertial modes
(see below).
At $\eta=0$, the coefficients $\kappa_0$ for the $r^o$- and $r^s$-modes are both equal to 
$2m/[l^\prime(l^\prime+1)]$, which
suggests that in the lowest order of $\Omega$ the two $r$-modes are 
degenerate at $\eta=0$.
In Figure 3, the toroidal components $iT_m$ and $iT_{m+2}$ 
of the $r^o$- and $r^s$-modes of $l^\prime=m=2$ at
$\bar\Omega=0.01$ are given versus $a/R$ for the case of $\eta=0$.
The amplitudes of $iT_{m+2}$ are much smaller than those of $iT_m$ 
both for the $r^o$- and $r^s$-modes.
This figure also shows that $|iT_m^p-iT_m^n|\not\ll 1$ in the core
for the $r^o$-mode at $\eta=0$, for which we find that $\pmb{w}^\prime\propto\Omega$.

Inertial modes in the superfluid core are also split into 
ordinary and superfluid inertial modes, which we call $i^o$- and $i^s$-modes.
In Table 2, we have given $\kappa_0$ for $i^o$- and $i^s$-modes for $m=2$ and $\eta=0$, and
see, e.g., Lockitch \& Friedman (1999) and Yoshida \& Lee (2000a) for 
the classification scheme employed here for inertial modes.
For given $m$ and $l_0-|m|$,
we find at $\eta=0$ pairs of $i^o$- and $i^s$-modes that have close values of $\kappa_0$,
and the number of the pairs is equal to $l_0-|m|$.
As an example, 
for the case of $m=2$ and $\eta=0$, the eigenfunctions $iT_{l^\prime}$
are shown for the $i^o$- and $i^s$-modes 
of $\kappa_0=0.5180$ and $\kappa_0=0.5077$ in Figure 4, 
for those of $\kappa_0=0.4215$ and $\kappa_0=0.4060$ in Figure 5, and
for those of $\kappa_0=1.1046$ and $\kappa_0=1.1134$ in Figure 6.
The inertial modes in Figure 4 belong to $l_0-|m|=3$ and those in Figures 5 and 6 to
$l_0-|m|=5$.
Note that
the $r^o$- and $r^s$-modes belong to $l_0-|m|=1$ (e.g., Yoshida \& Lee 2000a).
It is generally observed for inertial modes with long radial wavelengths
that the two superfluids co-move in the core
for the $i^o$-modes, and they counter-move for the $i^s$-modes.
The coefficient $\kappa_0$ for the $i^o$-modes only weakly
depends on the entrainment parameter $\eta$ (see Figure 7).
The coefficient $\kappa_0$ for the $i^s$-modes, on the other hand, 
increases approximately linearly as $\eta$ is increased from $\eta=0$ (see Figure 8).
See Yoshida \& Lee (2002b) for extended discussions on inertial modes in the superfluid core.

The $r^s$-modes ($r^o$-modes) experience mode crossings with $i^o$-modes ($i^s$-modes)
as the parameter $\eta$ is increased from $\eta=0$.
For the case of $m=2$ and $\bar\Omega=0.01$, Figure 7 illustrates an avoided crossing
between the $r^s$-mode of $l^\prime=m$ and the $i^o$-mode that tends to 
$\kappa_0=1.1046$ as $\eta\rightarrow0$.
In this figure, the dashed line is given by equation (49).
For mode crossings between the $r^o$-mode and $i^s$-modes, on the other hand,
it is quite difficult to numerically discern whether the mode crossings result in
avoided crossing or degeneracy of the mode frequencies at the crossing point.
Most prominent among such mode crossings of the $r^o$-mode of $l^\prime=m=2$
are those with the $i^s$-modes that tend to 
$\kappa_0=0.5077$ and $\kappa_0=0.4060$ as $\eta\rightarrow 0$.
For the case of $m=2$ and $\bar\Omega=0.01$,
the two mode crossings which occur at 
$\eta\approx0.0230$ and 0.0484 are shown in Figure 8, where the solid lines and the dashed line 
are for the $i^s$-modes and the $r^o$-mode, respectively.
We note that the ratio $\omega/\Omega\approx \kappa_0$ for the $r^o$-mode is almost constant, 
equal to $2m/[l^\prime(l^\prime+1)]=0.6667$ for $l^\prime=m=2$, while the ratios
$\omega/\Omega$ for the $i^s$-modes increase approximately linearly with increasing $\eta$.
Figure 9 shows the differences $iT_m^p-iT_m^n$ versus $a/R$ for the $l^\prime=m=2$ $r^o$-modes
at $\eta=0.023$ (panel a) and 0.0484 (panel b), and Figure 10 the eigenfunctions $iT_{l^\prime}$
for the $i^s$-modes of $m=2$ at $\eta=0.023$ (panel a) and at $\eta=0.0484$ (panel b), 
where we have assumed $\bar\Omega=0.01$ and the normalization $\max(|iT_m|)=1$.
The amplitudes of the differences at the mode crossing points are much larger than 
amplitudes of the difference off the crossing points (see, e.g., Figure 2).
The resemblance between the $i^s$-modes of $l_0-|m|=3$ at $\eta=0$ (Figure 4b)
and at $\eta=0.023$ (Figure 10a) and between
the $i^s$-modes of $l_0-|m|=5$ at $\eta=0$ (Figure 5b) and $\eta=0.0484$ (Figure 10b)
is obvious.

>From Table 1, we find that
the growth timescales $\tau^0_{GJ}$ and $\tau^0_{GD}$ for
the $r^o$-modes of $l^\prime=m=2$ are almost the same as those for the $l^\prime=m=2$ $r$-modes 
of the $N=1$ polytropic model without superfluidity in the core 
(see, e.g., Yoshida \& Lee 2000a).
On the other hand, the growth timescales $\tau^0_{GJ}$ and $\tau^0_{GD}$ for the
$r^s$-modes are much longer than those of the $r^o$-modes, and
the instability caused by the $r^s$-modes
is much weaker than the instability by the $r^o$-modes.
This is because the amount of gravitational wave radiation 
emitted from the $r^s$-modes is much smaller than that from the $r^o$-modes
since $(\rho_piT_m^p+\rho_niT_m^n)/\rho\sim 0$ and 
$\rho^\prime\sim 0$ for the $r^s$-modes.
It is interesting to note that $\tau^0_{GJ}$ of the $r^s$-modes at $\eta\not=0$ is 
by several orders of magnitude longer than that at $\eta=0$.
Figure 11 shows, as a function of $a/R$, $\sqrt{|\omega\sigma^{2l+1}|N_l/2E}J_{mm}^\prime(a)$ for
the $r^s$-modes at $\eta=0$ (dashed line) and at $\eta=0.04$ (solid line), where
$J_{mm}^\prime(a)=\int_0^adadJ_{mm}^\prime/da$.
As found for the case of $\eta=0.04$, the negative contributions to
$J_{mm}^\prime(a=R)$ in the envelope almost completely cancel out the positive contributions 
in the core, which leads to extremely long growth timescales $\tau^0_{GJ}$
of the $r^s$-modes at $\eta\not=0$.

Figure 12 illustrates the dependence of $-\tau^0_{MF}$ on the entrainment parameter $\eta$ for
the $l^\prime=m=2$ $r^o$-mode, and we find that
$-\tau^0_{MF}$ has prominent and deep minimums at $\eta\approx0.230$ and $0.484$.,
which is consistent with the result by Lindblom \& Mendell (2000).
Note that we have assumed $\rho_s=2.8\times 10^{14}$g/cm$^3$ for the interface between the
core and the envelope.
These prominent minimums result from the mode crossings with the long radial wavelength
$i^s$-modes that tend to $\kappa_0=0.5077$ and $\kappa_0=0.4060$ as $\eta\rightarrow 0$ 
(Figure 8; see also Andersson \& Comer 20001).
Mode crossings of the $r^o$-mode with $i^s$-modes that have much shorter radial wavelengths result
in narrow and shallow minimums of $-\tau^0_{MF}$, as found for $\eta\gtsim 0.05$.
These minor dips were not found by Lindblom \& Mendell (2000).
We guess that the $r$-modes calculated by Lindblom \& Mendell (2000) 
are less strongly affected by inertial modes associated with large $l_0-|m|$ having
short radial wavelengths since they ignored terms higher than $\Omega^4$ in their 
perturbative method.
Comparing our results with those by Lindblom \& Mendell (2000), we find that
the damping timescale $-\tau^0_{MF}$ at a given $\eta$ is about by an order of magnitude
shorter than that they obtained, and that our values of $\eta\approx0.0230$ and 0.0484
for the local minimums of $-\tau^0_{MF}$ are slightly larger than their $critical$ values 
0.02294 and 0.04817.
From Table 1 we also find that the value of $\kappa_2$ found here for the $r^o$-modes
is about by 2.5\% larger than the value they found.
We guess that these differences between the
two calculations partly come from the differences in the way of evaluating the
derivatives of the thermodynamical quantities that appear in the oscillation equations.
For example, Lindblom \& Mendell (2000) used
$\rho^\prime=(\partial \rho/\partial p)_\beta p^\prime+(\partial\rho/\partial\beta)_p\beta^\prime$,
for which they assumed $(\partial \rho/\partial p)_\beta=(\rho/2p)$ from
the polytropic relation $p=K\rho^2$ and employed
a fitting formula to $(\partial\rho/\partial\beta)_p$ obtained from
Akmal et al's equation of state (1998), where $\beta=\tilde{\mu}-\mu_n$.
In this paper, on the other hand, we used
$\rho^\prime=({\cal Q}_{11}+\zeta{\cal Q}_{21})\mu_n^\prime
+({\cal Q}_{12}+\zeta{\cal Q}_{22})\mu_p^\prime$,
and the coefficients ${\cal Q}_{ij}$ were all calculated by using equations (23) to (29)
with Akmal et al's equation of state (1998).

The eigenfunction $\delta\beta(r,\theta,\phi)=\delta\beta_0(r,\theta,\phi)
+(4/3)\delta\beta_2(r,\theta,\phi)\bar\Omega^2+O(\bar\Omega^4)$ in Lindblom \& Mendell (2000) may be
given in terms of the eigenfunctions $y^{p}_{2,k}$ and $y^{n}_{2,k}$ as
\begin{equation}
\delta \beta(r,\theta,\phi)=(M_a/M)(R/a)\bar\Omega^{-2}
\sum_{k\ge 1}\Delta y_{2,k}(a)Y_{l_km}(\theta,\phi),
\end{equation}
where $r=a[1+\epsilon(a,\theta)]$, and $\Delta y_{2,k}(a)=y_{2,k}^p(a)-y_{2,k}^n(a)$ 
and see Appendix for definition of the functions $y_{2,k}^{p(n)}(a)$.
Assuming $\delta\beta_0=0$, we obtain
\begin{equation}
\delta \beta_2(r,\theta,\phi)\approx 0.75\times(M_r/M)(R/r)\bar\Omega^{-4}
\sum_{k\ge 1}\Delta y_{2,k}(r)f_{l_km}P_{l_k}^m(\cos\theta)e^{im\phi}
\equiv\sum_{k\ge1}\delta\beta_{2,k}P_{l_k}^m(\cos\theta)e^{im\phi},
\end{equation}
where the mean radial distance $a$ have been replaced by the radial distance $r$, and
the factor $f_{lm}$ is defined by the relation 
$Y_l^m(\theta,\phi)=f_{lm}P^m_l(\cos\theta)e^{im\phi}$, and $l_k=|m|+2k-1$.
In Figure 13, we plot the functions $\delta\beta_{2,k}(r)$ 
for the $r^o$-mode of $l^\prime=m=2$ at $\eta=0.04$, applying amplitude normalization
given by $y_{2,k=1}(R)=f_{m+1,m}\bar\Omega^2$ at the surface, 
where the solid, dotted, and dashed lines
are for $\delta\beta_{2,1}$, $100\times\delta\beta_{2,2}$, and $100\times\delta\beta_{2,3}$, 
respectively.
Since $|\delta\beta_{2,1}(r)|>>|\delta\beta_{2,k}(r)|$ for $k\ge2$, the $\theta$ depenence 
of the function 
$\delta\beta_2(r,\theta,\phi)$ is well represented by a single associated Legendre function 
$P^m_{m+1}(\cos\theta)$, and 
is not necessarily the same $\theta$ dependence of the function $\delta\beta_2$ 
found by Lindblom \& Mendell (2000).
The amplitude of the function $\delta\beta_{2,1}(r)$ is about by a factor of $3$ larger than
the amplitude of $\delta\beta_2$ calculated by Lindblom \& Mendell (2000), which is
consistent with the result that $\tau_{MF}^0$ in this paper 
is about by an order of magnitude
shorter than $\tau_{MF}^0$ they obtained.

\section{Discussion}

If we employ a set of the dependent variables defined as (see Lindblom \& Mendell 2000)
\begin{equation}
\pmb{\xi}={\rho_n\pmb{v}_n^\prime+\rho_p\pmb{v}_p^\prime\over i\omega\hat\rho}, \quad
\pmb{\xi}^w={\pmb{w}^\prime\over i\omega}, \quad 
U={p^\prime\over\hat\rho}+\Psi^\prime, \quad
\beta^\prime=\tilde{\mu}^\prime-\mu^\prime_n,
\end{equation}
%
%={\rho_n\pmb{\xi}^n+\rho_p\pmb{\xi}^p\over\hat\rho}
%={\rho_n\rho_p\over\tilde\rho^2}(\pmb{\xi}^p-\pmb{\xi}^n)
%
where $\hat\rho=\rho_n+\rho_p$,
the continuity equations (16) and (17) and the velocity equations (18) and (19)
are rewritten as
\begin{equation}
\rho_n^\prime+\rho_p^\prime+\nabla\cdot\left(\hat\rho\pmb{\xi}\right)=0,
\end{equation}
\begin{equation}
{\rho_p^\prime\over\rho_p}-{\rho_n^\prime\over\rho_n}
+\pmb{\xi}\cdot\nabla\ln{\rho_p\over\rho_n}
+{\tilde\rho^2\over\rho_n\rho_p}\left(
\pmb{\xi}^w\cdot\nabla\ln{\tilde\rho^2\over\hat\rho}+\nabla\cdot\pmb{\xi}^w\right)=0,
\end{equation}
\begin{equation}
\pmb{F}\left(\pmb{\xi}\right)
=-\nabla U
+{\rho_n\rho_p\over\hat\rho^2}\left(\nabla\ln{\rho_p\over\rho_n}\right)\beta^\prime,
\end{equation}
\begin{equation}
\pmb{F}\left(\pmb{\xi}^w\right)+i\omega{\rho_{np}\hat\rho\over\rho_n\rho_p}
\pmb{D}\left(\pmb{\xi}^w\right)=-\nabla\beta^\prime,
\end{equation}
and the linearized Poisson equation remains the same:
\begin{equation}
\nabla^2\Psi^\prime=4\pi G\left(\rho_n^\prime+\rho_p^\prime\right),
\end{equation}
where $\pmb{D}(\pmb{\xi})=\pmb{\xi}\times(\nabla\times\pmb{v}_0)$, and
the terms of order of $m_e/m_p$ have been ignored for simplicity.
The term proportional to $\pmb{D}\left(\pmb{\xi}^w\right)$ in equation (56) 
represents the drag force between the two superfluids.
In a perturbative treatment of the $r$-modes,
we may expand the eigenfunctions and eigenfrequencies in terms of $\Omega$ as
\begin{equation}
\pmb{\xi}=\pmb{\xi}_0+\pmb{\xi}_2\Omega^2+O(\Omega^4), \quad
\pmb{\xi}^w=\pmb{\xi}_0^w+\pmb{\xi}_2^w\Omega^2+O(\Omega^4), 
\end{equation}
\begin{equation}
U=U_2\Omega^2+U_4\Omega^4+O(\Omega^6), \quad
\beta^\prime=\beta_2^\prime\Omega^2+\beta_4^\prime\Omega^4+O(\Omega^6), 
\end{equation}
and
\begin{equation}
\omega=\kappa_0\Omega+\kappa_2\Omega^3+O(\Omega^5), \quad
\sigma=s_0\Omega+s_2\Omega^3+O(\Omega^5),
\end{equation}
where $|\pmb{\xi}_0|\sim |iT_m|\sim O(1)$.
Note that the quantities like $\nabla\cdot\pmb{\xi}$,
$\pmb{\xi}\cdot\nabla\rho$, $\rho_n^\prime$, 
$\rho_p^\prime$, and $\Psi^\prime$ are of order of $\Omega^2$ in the
lowest order for the $r$-modes.
For the expansions given above, we have
\begin{equation}
\pmb{F}(\pmb{\xi})=\pmb{F}_2(\pmb{\xi}_0,\kappa_0,s_0)\Omega^2
+\pmb{F}_4(\pmb{\xi}_0,\pmb{\xi}_2,\kappa_0,\kappa_2,s_0,s_2)\Omega^4
+O(\Omega^6), 
\end{equation}
and we may not need to give the full expressions of $\pmb{F}_2$ and $\pmb{F}_4$ here.
Equations (55) and (56) suggest the existence of two families of $r$-modes, that is,
$r$-modes, which are governed by equation (55) and have
$\kappa_0=2m/[l^\prime(l^\prime+1)]$, and $r$-modes, which are governed by equation (56)
and have $\kappa_0\not=2m/[l^\prime(l^\prime+1)]$.
If we assume $\pmb{F}_2(\pmb{\xi}_0)=-\nabla U_2$ in equation (55), we obtain 
the lowest order $r$-mode solution given by $iT_m\propto a^{|m|-1}$ and
$\kappa_0=2m/[l^\prime(l^\prime+1)]$, which leads to $\beta_2^\prime=0$ from equation (55),
and to $\pmb{\xi}^w_0=0$ from equation (56).
This suggests the existence of the $r$-mode solutions, for which 
$\kappa_0=2m/[l^\prime(l^\prime+1)]$, and $\pmb{\xi}_0\not=0$ and $U_2\not=0$, but
$\pmb{\xi}^w_0=0$ and $\beta^\prime_2=0$.
The solutions of this kind correspond to 
the $r$-modes Lindblom \& Mendell (2000) obtained in their perturbative treatment, and
to the $r^o$-modes found in this paper.
On the other hand, because of the drag force, i.e.,
the term proportional to $\pmb{D}(\pmb{\xi}^w)$,
$\kappa_0$ of the $r$-modes governed by equation (56)
is not necessarily equal to $2m/[l^\prime(l^\prime+1)]$.
For these solutions we may expect 
$\pmb{\xi}^w_0\not=0$ and $\beta_2^\prime\not=0$ from equation (56), and 
$\pmb{\xi}_0\not=0$ and $U_2\not=0$ from equation (55).
The solutions of this kind
correspond to the $r^s$-modes found in this paper (see also Andersson \& Comer 2001).

To show the importance of the drag force for the existence of the $r$-mode solutions
with the scaling $\pmb{w}^\prime\propto\Omega^3$, we calculate
the $r^o$- and $r^s$-modes by ignoring the drag force terms proportional to $\pmb{D}(\pmb{\xi})$ 
on the right hand side of the velocity equations (18) and (19).
The results are summarized in Table 3, in which
the coefficients $\kappa_0$, $\kappa_2$, and $\tau^0_i$'s are tabulated
for the $r^o$- and $r^s$-modes of $l^\prime=m=2$ for the case of $\eta=0.04$.
When we ignore the drag force terms, we can not find the $r^o$-modes with the
scaling of $\pmb{w}^\prime\propto\Omega^3$ and
the $r$-modes with the scaling $\pmb{w}^\prime\propto\Omega$ are the only $r$-mode
solutions we can find, and hence
the exponent $\gamma$ to define $\tau^0_{MF}$ in the table
is equal to $0.5$ for both the $r^o$- and $r^s$-modes.
In Figure 14, the toroidal components $iT_m$ and $iT_{m+2}$ of the two $r$-modes
are given for $\eta=0.04$, where we have assumed $\bar\Omega=0.01$ and 
the normalization $\max(|iT_m|)=1$.
The amplitudes of $iT_{m+2}$ are completely negligible compared to those of $iT_m$.
In the absence of the drag force,
$\kappa_0$'s for the two $r$-modes are both equal to
$2m/[l^\prime(l^\prime+1)]$, independent of $\eta$, which means that the frequencies of 
the two $r$-modes are degenerate in the lowest order of $\Omega$.
These results suggest that
the $r$-mode solutions with $\pmb{w}^\prime\propto\Omega^3$ can be found
only when the frequencies of the $r$-modes are not degenerate in the lowest order in $\Omega$.
In this sense, it is reasonable to find $\pmb{w}^\prime\propto\Omega$ for the $r^o$-modes
at $\eta=0$ since $\kappa_0$'s of the $r^o$- and $r^s$-modes are equal to each other 
at $\eta=0$, as shown by Table 1.

\section{Conclusion}

In this paper we have discussed the modal properties of the $r$-modes of neutron stars with
the core filled with neutron and proton superfluids.
We numerically find that
the $r$-modes of rotating neutron stars with the superfluid core are split into ordinary $r^o$-modes 
and superfluid $r^s$-modes, and that
the two superfluids in the core flow together for the $r^o$-modes and they counter move
for the $r^s$-modes.
These findings are consistent with earlier suggestions made analytically by Andersoon \& Comer (2001).
We also find that,
although $\kappa_0$ of the $r^o$-modes is numerically equal to $2m/[l^\prime(l^\prime+1)]$,
almost independent of the entrainment parameter $\eta$,
$\kappa_0$ of the $r^s$-modes approximately linearly increases from
$2m/[l^\prime(l^\prime+1)]$ as $\eta$ is increased from $\eta=0$.
The $r^o$-modes have the scaling 
of $\pmb{w}^\prime\propto\Omega^3$ and
are the same $r$-modes discussed by Lindblom \& Mendell (2000), while 
the $r^s$-modes have the scaling of $\pmb{w}^\prime\propto\Omega$.
The instability caused by the $r^s$-modes is found much weaker than the instability 
by the $r^o$-mode and will be
easily stabilized by various dissipation processes in the star.

We have confirmed that,
except in a few narrow regions of $\eta$ around the prominent local minimums of $\tau^0_{MF}$,  
the dissipation due to mutual friction in the superfluid core is ineffective 
to stabilize the instability by the $r^o$-modes, as first shown by
Lindblom \& Mendell (2000).
We have shown that these prominent local minimums of $-\tau^0_{MF}$ are caused by
mode crossings between the $r^o$-mode and 
the superfluid inertial $i^s$-modes
with long radial wavelengths comparable to those of the $r^o$-mode 
(see also Andersson \& Comer 2001).
Since mutual friction is almost always very strong for the $r^s$-modes and the instability
by the $r^s$-modes itself is quite weak,
only the instability by the $r^o$-modes will be of direct observational importance, e.g., as 
mechanisms that generate
gravitational waves from normal modes of rotating neutron stars and/or
cause the clustering of spin frequencies around 300Hz for neutron stars in
LMXB's, unless there exist other strong damping mechanisms for the $r$-modes 
(see Jones 2001a,b;
Lindblom \& Owen 2002; Haensel, Levenfish \& Yakovlev 2002).

In this paper, we have made a brief report on a numerical result of inertial modes in
the superfluid core.
We find that inertial modes in superfluid neutron stars are also split into ordinary 
inertial $i^o$-modes and superfluid inertial $i^s$-modes.
It is generally observed for inertial modes with long radial wavelengths
that the two superfluids co-move in the core
for the $i^o$-modes, and they counter-move for the $i^s$-modes
(see Yoshida \& Lee 2002b for
more complete discussions on inertial modes in the superfluid core).
Inertial modes are expected to work as a mechanism limiting 
the amplitude growth of the $r$-modes (Morsink 2002).

There are many interesting and challenging problems we have to deal with
before we can conclude definitely about the neutron star oscillations.
Confronting their cooling calculations of neutron stars
with observational data for the surface temperature of several neutron
stars, Kaminker, Haensel, \& Yakovlev (2001), and 
Kaminker, Yakovlev, \& Gnedin (2002) suggested that neutron superfluidity
in the core of middle-aged neutron stars should be weak in the sense that
the critical temperature $T_c$ is less than $10^8$K. 
If this is the case, there exist no neutron superfluids in the core of
temperatures $T>10^8$K.
If neutrons in the core are in normal state, modal properties of low frequency oscillations 
propagating in the core will be different
from those for the core filled with neutron and proton superfluids,
since buoyant force in the core, produced by thermal and/or chemical stratification 
(Reisenegger \& Goldreich 1992), comes into play.
The normal neutron fluid core with stratification will support $g$-mode propagation, 
and the buoyant force in it
will affect inertial modes in slowly rotating neutron stars.
The nodeless $r$-modes of $l^\prime=m$, however, will remain almost the same even 
in the presence of buoyant force (see Yoshida \& Lee 2000b), and 
damping due to mutual friction in the core will remain weak for the $r^o$-modes,
for which neutron and proton fluids co-move.
Let us point out a possibility of using neutron star binaries as a probe 
to investigate existence or non-existence of neutron superfludity in the core,
since tidally excited low frequency modes will be $g$-modes in the normal fluid core but
inertial modes in the superfluid core, and the different tidal responses 
will result in different binary evolutions.
The presence of a solid crust and a magnetic field in neutron star interior 
is another factor that makes complicated the problems of global oscillations of the stars.
The solid crust is not a rigid body and supports its own normal modes 
(see, e.g., McDermott et al 1988, Lee \& Strohmayer 1996).
For example, there exist torsional modes propagating in the crust, which
can be resonantly coupled with the $r$-modes in the normal fluid core
(Yoshida \& Lee 2001; Levin \& Ushomirsky 2001).
Quite recently, using a local analysis of perturbations, Kinney \& Mendell (2002) 
have suggested that
no $r$-mode solution to the magnetohydrodynamic equations (e.g., Mendell 1998)
exists in the superfluid core when
both the neutron and proton vortices are pinned to the solid crust.
This may suggest that a careful formulation of boundary conditions at the core-crust interfaces
will be necessary to obtain reliable solutions to global oscillations of neutron stars,
particularly when a magnetic field is essential for the oscillations.

\appendix

\section{Oscillation Equations in the Superfluid Core}

For the oscillation equations in the superfluid core,
we employ vectors $\pmb{y}_1^n$, $\pmb{y}_2^n$, $\pmb{y}_1^p$, $\pmb{y}_2^p$,
$\pmb{y}_3$, and $\pmb{y}_4$, whose components are given by
\begin{equation}
y_{1,k}^n=S_l^n, \quad y_{2,k}^n={\mu_{n,l}^\prime+\Psi^\prime_l\over ga}, \quad
y_{1,k}^p=S_l^p, \quad y_{2,k}^p={\tilde\mu_{l}^\prime+\zeta\Psi^\prime_l\over ga}, \quad
y_{3,k}={\Psi^\prime_l\over ga}, \quad y_{4,k}={1\over g}{d\Psi^\prime_l\over da},
\end{equation}
where $l=|m|+2(k-1)$ for even modes and $l=|m|+2k-1$ for odd modes, and
$k=1,~2,~3, \cdots$.
We also introduce vectors $\pmb{h}^n$, $\pmb{h}^p$, $i\pmb{t}^n$, and $i\pmb{t}^p$,
the components of which are
\begin{equation}
h_k^n=H_l^n, \quad it_k^n=iT_{l^\prime}^n, \quad 
h_k^p=H_l^p, \quad it_k^p=iT_{l^\prime}^p, 
\end{equation}
where $l^\prime=l+1$ for even modes and $l^\prime=l-1$ for odd modes.
Using these vectors,
the perturbed continuity equations (16) and (17) are written as
\begin{eqnarray}
a{d\pmb{y}_1^n\over da}&=&\left(-{d\ln\rho_n\over d\ln a}-3-a{d\chi_3(\alpha)\over da}
-a{d\chi_3(\beta)\over da}A_0\right)\pmb{y}_1^n-{ga\over\rho_n}{\cal Q}_{11}\pmb{y}_2^n
-{ga\over\rho_n}{\cal Q}_{12}\pmb{y}_2^p
+{ga\over\rho_n}\left({\cal Q}_{11}+\zeta {\cal Q}_{12}\right)\pmb{y}_3 \nonumber  \\
&+& \left(\Lambda_0+3\chi_3(\beta)B_0\right)\pmb{h}^n
+3m\chi_3(\beta)Q_0i\pmb{t}^n, 
\end{eqnarray}
\begin{eqnarray}
a{d\pmb{y}_1^p\over da}&=&\left(-{d\ln\rho_p\over d\ln a}-3-a{d\chi_3(\alpha)\over da}
-a{d\chi_3(\beta)\over da}A_0\right)\pmb{y}_1^p-{ga\over\rho_p}{\cal Q}_{21}\pmb{y}_2^n
-{ga\over\rho_p}{\cal Q}_{22}\pmb{y}_2^p
+{ga\over\rho_p}\left({\cal Q}_{21}+\zeta {\cal Q}_{22}\right)\pmb{y}_3 \nonumber  \\
&+& \left(\Lambda_0+3\chi_3(\beta)B_0\right)\pmb{h}^p
+3m\chi_3(\beta)Q_0i\pmb{t}^p.
\end{eqnarray}
The radial components of equations (18) and (19) are reduced to
\begin{eqnarray}
a{d\pmb{y}_2^n\over da}&=&c_1\bar\omega^2{\rho_{11}\over\rho_n}E_0\pmb{y}_1^n+(1-U)\pmb{y}_2^n
+c_1\bar\omega^2{\rho_{12}\over\rho_n}E_0\pmb{y}_1^p \nonumber \\
&-&c_1\bar\omega^2\left({\rho_{11}\over\rho_n}3\beta B_0+m\nu E_1\right)\pmb{h}^n
-c_1\bar\omega^2{\rho_{12}\over\rho_n}3\beta B_0\pmb{h}^p  \nonumber \\
&-&c_1\bar\omega^2\left({\rho_{11}\over\rho_n}3m\beta Q_0+\nu E_1C_0\right)i\pmb{t}^n
-c_1\bar\omega^2{\rho_{12}\over\rho_n}3m\beta Q_0i\pmb{t}^p, 
\end{eqnarray}
\begin{eqnarray}
a{d\pmb{y}_2^p\over da}&=&c_1\bar\omega^2\left({\rho_{22}\over\rho_p}+{m_e\over m_p}\right)
E_0\pmb{y}_1^p+(1-U)\pmb{y}_2^p
+c_1\bar\omega^2{\rho_{21}\over\rho_p}E_0\pmb{y}_1^n \nonumber \\
&-&c_1\bar\omega^2\left[\left({\rho_{22}\over\rho_p}+{m_e\over m_p}\right)3\beta B_0
+m\nu\zeta E_1\right]\pmb{h}^p
-c_1\bar\omega^2{\rho_{21}\over\rho_p}3\beta B_0\pmb{h}^n \nonumber \\
&-&c_1\bar\omega^2\left[\left({\rho_{22}\over\rho_p}+{m_e\over m_p}\right)3m\beta Q_0
+\nu\zeta E_1C_0\right]i\pmb{t}^p
-c_1\bar\omega^2{\rho_{21}\over\rho_p}3m\beta Q_0i\pmb{t}^n.
\end{eqnarray}
The perturbed Poisson equation (21) is reduced to
\begin{equation}
a{d\pmb{y}_3\over da}=\left(1-U\right)\pmb{y}_3+\pmb{y}_4,
\end{equation}
\begin{eqnarray}
a{d\pmb{y}_4\over da}&= & 4\pi G a^2\left({\cal Q}_{11}+\zeta{\cal Q}_{21}\right)I^{-1}\pmb{y}_2^n
+4\pi G a^2\left({\cal Q}_{12}+\zeta{\cal Q}_{22}\right)I^{-1}\pmb{y}_2^p \nonumber \\
&+&I^{-1}\left[\Lambda_0-4\pi G a^2\left({\cal Q}_{11}
+\zeta{\cal Q}_{21}+\zeta({\cal Q}_{12}+\zeta{\cal Q}_{22})\right)\pmb{1}
-2\left(\alpha\pmb{1}+\beta A_0\right)\Lambda_0\right]\pmb{y}_3  \nonumber \\
&+& I^{-1}\left[(-U+\chi_0(\alpha))I
+\chi_0(\beta)A_0-6\beta (A_0+B_0)\right]\pmb{y}_4.
\end{eqnarray}
Algebraic equations that determine the relations between the variables $\pmb{h}$, $i\pmb{t}$,
$\pmb{y}_1$, and $\pmb{y}_2$ are derived by making use of the horizontal components of
the velocity equations (18) and (19), and they are
\begin{equation}
\tilde L_0\pmb{h}^n-\tilde M_1i\pmb{t}^n
-f_nmF_0(\pmb{h}^p-\pmb{h}^n)-f_n\tilde M_1^+(i\pmb{t}^p-i\pmb{t}^n)
= \tilde J\pmb{y}_1^n +f_n \tilde J^+(\pmb{y}_1^p-\pmb{y}_1^n)
+b_{11}{\pmb{y}_2^n\over c_1\bar\omega^2}+b_{12}{\pmb{y}_2^p\over c_1\bar\omega^2},
\end{equation}
\begin{equation}
\tilde L_0\pmb{h}^p-\tilde M_1i\pmb{t}^p
+ f_pmF_0(\pmb{h}^p-\pmb{h}^n)+f_p\tilde M_1^+(i\pmb{t}^p-i\pmb{t}^n)
= \tilde J\pmb{y}_1^p-f_p \tilde J^+(\pmb{y}_1^p-\pmb{y}_1^n)
+b_{21}{\pmb{y}_2^n\over c_1\bar\omega^2}+b_{22}{\pmb{y}_2^p\over c_1\bar\omega^2},
\end{equation}
\begin{equation}
-\tilde M_0\pmb{h}^n+\tilde L_1i\pmb{t}^n
-f_n\tilde M_0^+(\pmb{h}^p-\pmb{h}^n)
-f_nmF_1(i\pmb{t}^p-i\pmb{t}^n)
=-\tilde K\pmb{y}_1^n-f_n\tilde K^+(\pmb{y}_1^p-\pmb{y}_1^n),
\end{equation}
\begin{equation}
-\tilde M_0\pmb{h}^p+\tilde L_1i\pmb{t}^p
+f_p\tilde M_0^+ (\pmb{h}^p-\pmb{h}^n)
+f_pmF_1(i\pmb{t}^p-i\pmb{t}^n)
=-\tilde K\pmb{y}_1^p+f_p\tilde K^+(\pmb{y}_1^p-\pmb{y}_1^n),
\end{equation}
where
\begin{equation}
f_n={\rho_{np}\over\rho_n}{\zeta\over\tilde\zeta}, \quad
f_p={\rho_{np}\over\rho_p}{1\over\tilde\zeta},
\end{equation}
\begin{equation}
\zeta=1+{m_e\over m_p}, \quad
\tilde\zeta=1+{m_e\over m_p}{\rho_{pp}\over\rho_p},
\end{equation}
and
\begin{equation}
\left(\matrix{b_{11}&b_{12}\cr b_{21}&b_{22}\cr}\right)=
\left(\matrix{\rho_{11}/\rho_n&\rho_{12}/\rho_n\cr
\rho_{21}/\rho_p&(\rho_{22}/\rho_p)+(m_e/m_p)\cr}
\right)^{-1}.
\end{equation}
The functions $\chi_1(\alpha)$, $\chi_2(\alpha)$, $\chi_3(\alpha)$, and $\chi_0(\alpha)$ 
are defined as
\begin{equation}
\chi_1(\alpha)=\alpha+a{d\alpha\over da}, \quad
\chi_2(\alpha)=2\alpha+a{d\alpha\over da}, \quad
\chi_3(\alpha)=3\alpha+a{d\alpha\over da},
\end{equation}
and
\begin{equation}
\chi_0(\alpha)=-a{d\alpha\over da}+a{d\over da}\left(a{d\alpha\over da}\right).
\end{equation}
The matrices $E_0$, $E_1$, $F_0$, $F_1$, $I$, $\tilde J$, $\tilde J^+$, $\tilde K$, $\tilde K^+$,
$\tilde L_0$, $\tilde L_0^+$, $\tilde L_1$, $\tilde L_1^+$, 
$\tilde M_0$, $\tilde M_0^+$, $\tilde M_1$, and $\tilde M_1^+$ are defined as
\begin{equation}
E_0=\left(1+2\chi_1(\alpha)\right)\pmb{1}+2\chi_1(\beta)A_0, \quad
E_1=(1+\chi_2(\alpha))\pmb{1}+\chi_2(\beta)A_0,
\end{equation}
\begin{equation}
F_0=\nu\Lambda_0^{-1}\left[(1+2\alpha+2\beta)\pmb{1}+12\beta A_0\right], \quad
F_1=\nu\Lambda_1^{-1}\left[(1+2\alpha+2\beta)\pmb{1}+12\beta A_1\right],
\end{equation}
\begin{equation}
I=\pmb{1}-2\chi_1(\alpha)\pmb{1}-2\chi_1(\beta)A_0,
\end{equation}
\begin{equation}
\tilde J=\tilde J^+-3\beta\Lambda_0^{-1}(2A_0+B_0), 
\end{equation}
\begin{equation}
\tilde J^+=m\nu\Lambda_0^{-1}\left[(1+\chi_2(\alpha))\pmb{1}+\chi_2(\beta)A_0\right],
\end{equation}
\begin{equation}
\tilde K =\tilde K^+-3m\beta\Lambda_1^{-1}Q_1,
\end{equation}
\begin{equation}
\tilde K^+
=\nu\Lambda_1^{-1}\left[((1+\chi_2(\alpha))\pmb{1}+\chi_2(\beta)A_1)C_1
+2((1+\chi_2(\alpha)-\chi_2(\beta))\pmb{1}+2\chi_2(\beta)A_1)Q_1\right],
\end{equation}
\begin{equation}
\tilde L_0=(1+2\alpha)\pmb{1}-mF_0+\beta\Lambda_0^{-1}(2A_0\Lambda_0+6B_0), 
\end{equation}
\begin{equation}
\tilde L_1=(1+2\alpha)\pmb{1}-mF_1+\beta\Lambda_1^{-1}(2A_1\Lambda_1+6B_1), 
\end{equation}
\begin{equation}
\tilde M_0=\tilde M_0^+-6m\beta\Lambda_1^{-1}Q_1, 
\end{equation}
\begin{equation}
\tilde M_0^+=\nu\Lambda_1^{-1}
[(1+2\alpha-2\beta)\pmb{1}+4\beta A_1]Q_1\Lambda_0+F_1C_1, \quad
\end{equation}
\begin{equation}
\tilde M_1=\tilde M_1^+-6m\beta\Lambda_0^{-1}Q_0, 
\end{equation}
\begin{equation}
\tilde M_1^+=\nu\Lambda_0^{-1}
[(1+2\alpha-2\beta)\pmb{1}+4\beta A_0]Q_0\Lambda_1+F_0C_0,
\end{equation}
where $\pmb{1}$ denotes the unit matrix,
and the matrices $A_0$, $A_1$, $B_0$, $B_1$, $C_0$, $C_1$,
$Q_0$, $Q_1$, $\Lambda_0$, and $\Lambda_1$ as well as the quantities $U$ and $c_1$
are defined in Yoshida \& Lee (2000a).

The oscillation equations in the superfluid core are given as a set of linear ordinary
differential equations for the variables $\pmb{y}_j$ with $j=1$ to 4, which
are obtained by eliminating the vectors
$\pmb{h}$ and $i\pmb{t}$ in equations (A3) to (A8) 
using the algebraic equations (A9) to (A12).

\newpage

\begin{deluxetable}{cccccccc}
\footnotesize
\tablecaption{$\kappa_0$, $\kappa_2$, and 
$\tau^0_i$'s for the $r^o$- and $r^s$-modes of $l^\prime=m=2$.}
\tablewidth{0pt}
\tablehead{ \colhead{Modes}&\colhead{$\eta$}& \colhead{$\kappa_0$} & \colhead{$\kappa_2$}
& \colhead{$\tau^0_{GJ}$(s)} & \colhead{$\tau^0_{GD}$(s)} &
\colhead{$\tau^0_{S}$(s)} & \colhead{$\tau^0_{MF}$(s)} }
\startdata
 $r^o$& 0 & 0.6667 & 0.408 & $3.31$ & $3.83\times 10^2$ & $-7.55\times10^7$ &  $-\infty$ \\
$r^s$&  & 0.6667 & 0.511 & $2.12\times10^4$ & $2.18\times 10^6$  & $-2.99 \times 10^6$ 
& $-\infty$ \\
$r^o$& 0.02 & 0.6667 & 0.408 & $3.31$ & $3.83\times 10^2$ & $-6.76\times10^7$ 
&  $-2.52\times10^2$ \\
$r^s$ & & 0.8532 & 0.652 & $4.76\times10^{8}$ & $1.21\times 10^6$  & $-1.99 \times 10^6$ 
& $-2.94\times 10^{-1}$ \\
 $r^o$&0.04 & 0.6667 & 0.408 & $3.31$ & $3.83\times 10^2$ & $-6.76\times10^7$ 
&  $-1.70\times10^3$ \\
 $r^s$& & 1.0408 & 0.787 & $7.89\times10^{8}$ & $2.55\times 10^6$  & $-1.14 \times 10^6$ 
& $-7.99\times 10^{-2}$ \\
 $r^o$&0.06 & 0.6667 & 0.408 & $3.31$ & $3.83\times 10^2$ & $-6.76\times10^7$ 
&  $-2.62\times10^3$ \\
 $r^s$ && 1.2296 & 0.821 & $3.64\times10^{9}$ & $1.03\times 10^7$  & $-4.67 \times 10^4$ 
& $-5.20\times 10^{-2}$ \\
\enddata
\label{coefficients0}
\end{deluxetable}

\begin{deluxetable}{ccccc}
\footnotesize
\tablecaption{$\kappa_0$ for the $i^o$- and $i^s$-modes for $m=2$ and $\eta=0$.}
\tablewidth{0pt}
\tablehead{ \colhead{$l_0-|m|$}& \colhead{$i^o$} & \colhead{$i^s$}}
\startdata
2 & 1.0978 & 1.1190 \\
 & -0.5683 & -0.5109 \\
3 & 1.3564 & 1.3869 \\
  & 0.5180 & 0.5077 \\
  & -1.0293 & -0.9723 \\
4 & 1.5191 & 1.5526 \\
 & 0.8639 & 0.8612 \\
 & -0.2738 & -0.2417\\
 & -1.2738 & -1.2238 \\
5 & 1.6273 & 1.6600 \\
  & 1.1046 & 1.1134 \\
  & 0.4215 & 0.4060 \\
  & -0.7028 & -0.6524 \\
 & -1.4340 & -1.3968 \\
\enddata
\label{coefficients0}
\end{deluxetable}

\begin{deluxetable}{ccccccc}
\footnotesize
\tablecaption{$\kappa_0$, $\kappa_2$, and 
$\tau^0_i$'s for the $r^o$- and $r^s$-modes of $l^\prime=m=2$ for $\eta=0.04$ without the drag force.}
\tablewidth{0pt}
\tablehead{ \colhead{Modes}& \colhead{$\kappa_0$} & \colhead{$\kappa_2$}
& \colhead{$\tau^0_{GJ}$(s)} & \colhead{$\tau^0_{GD}$(s)} &
\colhead{$\tau^0_{S}$(s)} & \colhead{$\tau^0_{MF}$(s)} }
\startdata
$r^o$ & 0.6667 & 0.408 & $3.31$ & $3.83\times 10^2$ & $-7.49\times10^7$ &  $-9.10\times10^2$ \\
$r^s$ & 0.6667 & 0.529 & $4.33\times10^4$ & $2.63\times 10^6$  & $-1.13 \times 10^6$ & $-7.30\times10^{-2}$ \\
\enddata
\label{coefficients0}
\end{deluxetable}

\begin{figure}
\epsscale{.5}
%\plotone{f1.eps}

\caption{Toroidal components $iT_m$ and $iT_{m+2}$ of the $r$-modes of
$l^\prime=m=2$ are plotted 
as a function of $a/R$ for $\bar\Omega=0.01$ and $\eta=0.04$, 
where the solid, dotted, dashed, and dash-dotted lines
are used to indicate, respectively, $iT_m^n$, $iT_{m+2}^n$, $iT_m^p$, and $iT_{m+2}^p$
in the superfluid core.
The amplitudes are normalized by the maximum value.
Panels (a) and (b) are for the $r^o$- and $r^s$-modes, respectively.}

\end{figure}

\begin{figure}
\epsscale{.5}
%\plotone{f2.eps}

\caption{Difference $iT_m^p-iT_m^n$ versus $a/R$ for the $l^\prime=m=2$ $r^o$-mode
for $\bar\Omega=0.01$ and $\eta=0.04$, where 
the amplitude normalization $\max(|iT_m|)=1$ have been used.}

\end{figure}

\begin{figure}
\epsscale{.5}
%\plotone{f3.eps}

\caption{Same as Figure 1 but for the case of $\eta=0$.}

\end{figure}

\begin{figure}
\epsscale{.5}
%\plotone{f4.eps}

\caption{Toroidal components $iT_m$ and $iT_{m+2}$ of inertial modes of
$m=2$ and $l_0-|m|=3$ are plotted 
as a function of $a/R$ for $\bar\Omega=0.01$ and $\eta=0$, 
where the solid, dotted, dashed, and dash-dotted lines
are used to indicate, respectively, $iT_m^n$, $iT_{m+2}^n$, $iT_m^p$, and $iT_{m+2}^p$
in the superfluid core.
The amplitudes are normalized by the maximum value.
Panels (a) and (b) are for the $i^o$-mode of $\kappa_0=0.5180$ and the $i^s$-mode of
$\kappa_0=0.5077$, respectively.}

\end{figure}

\begin{figure}
\epsscale{.5}
%\plotone{f5.eps}

\caption{Same as Figure 4 but for $l_0-|m|=5$.
Panels (a) and (b) are for the $i^o$-mode of $\kappa_0=0.4215$ and the $i^s$-mode of
$\kappa_0=0.4060$, respectively.}

\end{figure}

\begin{figure}
\epsscale{.5}
%\plotone{f6.eps}

\caption{Same as Figure 4 but for $l_0-|m|=5$.
Panels (a) and (b) are for the $i^o$-mode of $\kappa_0=1.1046$ and 
the $i^s$-mode of $\kappa_0=1.1134$, respectively.}

\end{figure}

\begin{figure}
\epsscale{.5}
%\plotone{f7.eps}

\caption{Avoided crossing as a function of $\eta$
between the $l^\prime=m$ $r^s$-mode and the $i^o$-mode that tends to $\kappa_0=1.1046$ as
$\eta\rightarrow0$, where $m=2$ and $\bar\Omega=0.01$ have been assumed.
The dashed line indicates $\kappa_0$ given by equation (49) for the $r^s$-mode.}

\end{figure}

\begin{figure}
\epsscale{.5}
%\plotone{f8.eps}

\caption{Mode crossings as a function of $\eta$
between the $l^\prime=m$ $r^o$-mode (dashed line) and the $i^s$-modes (solid lines)
that tend to $\kappa_0=0.5077$ and $\kappa_0=0.4060$ as $\eta\rightarrow0$, 
where $m=2$ and $\bar\Omega=0.01$ have been assumed.}

\end{figure}

\begin{figure}
\epsscale{.5}
%\plotone{f9.eps}

\caption{Differences $iT_m^p-iT_m^n$ for the $l^\prime=m=2$ $r^o$-mode at $\bar\Omega=0.01$
are plotted  as a function of $a/R$ for the cases of $\eta=0.023$ (panel a)
and $\eta=0.0484$ (panel b).
The amplitude normalization is given by $\max(|iT_m|)=1$.}

\end{figure}

\begin{figure}
\epsscale{.5}
%\plotone{f10.eps}

\caption{Toroidal components $iT_m$ and $iT_{m+2}$ of inertial modes of
$m=2$ are plotted as a function of $a/R$ for $\bar\Omega=0.01$, 
where the solid, dotted, dashed, and dash-dotted lines
are used to indicate, respectively, $iT_m^n$, $iT_{m+2}^n$, $iT_m^p$, and $iT_{m+2}^p$
in the superfluid core.
The amplitudes are normalized by the maximum value.
Panel (a) shows the $i^s$-mode of $l_0-|m|=3$ at
$\eta=0.023$, and panel (b) the $i^s$-mode of $l_0-|m|=5$ at $\eta=0.0484$}

\end{figure}

\begin{figure}
\epsscale{.5}
%\plotone{f11.eps}

\caption{ $NJ^\prime_{mm}(a)$ versus $a/R$ for the $l^\prime=m=2$ $r^s$-mode
for the cases of $\eta=0.04$ (solid line)
and $\eta=0$ (dashed line), where $N\equiv \sqrt{|\omega\sigma^{2l+1}|N_l/2E}$
and $J^\prime_{mm}(a)=\int_0^adadJ^\prime_{mm}/da$.}

\end{figure}

\begin{figure}
\epsscale{.5}
%\plotone{f12.eps}

\caption{ $-\tau^0_{MF}$ versus $\eta$ for the $l^\prime=m=2$
$r^o$-mode, where $\rho_s=2.8\times10^{14}$g/cm$^3$ has been used, and
$-\tau^0_{MF}$ is given in second.}

\end{figure}

\begin{figure}
\epsscale{.5}
%\plotone{f13.eps}

\caption{ Eigenfunction $\delta\beta_{2,k}$ as a function of $r/R$ for the $r^o$-mode of
$l^\prime=m=2$ for $\bar\Omega=0.01$ and $\eta=0.04$, where amplitude normalizaation has been given
by $y_{2,k=1}(r=R)=f_{m+1,m}\bar\Omega^2$. See text for definition of the factor $f_{lm}$ and 
the function $y_{2,k}$. The solid, dotted, and dashed lines are for
$\delta\beta_{2,1}$, $100\times \delta\beta_{2,2}$, and $100\times \delta\beta_{2,3}$, respectively.}

\end{figure}

\begin{figure}
\epsscale{.5}
%\plotone{f14.eps}

\caption{Same as Figure 1 but for that
we have ignored the drag force terms on the right hand side of the velocity equations
(18) and (19) to calculate the $r$-modes.}

\end{figure}

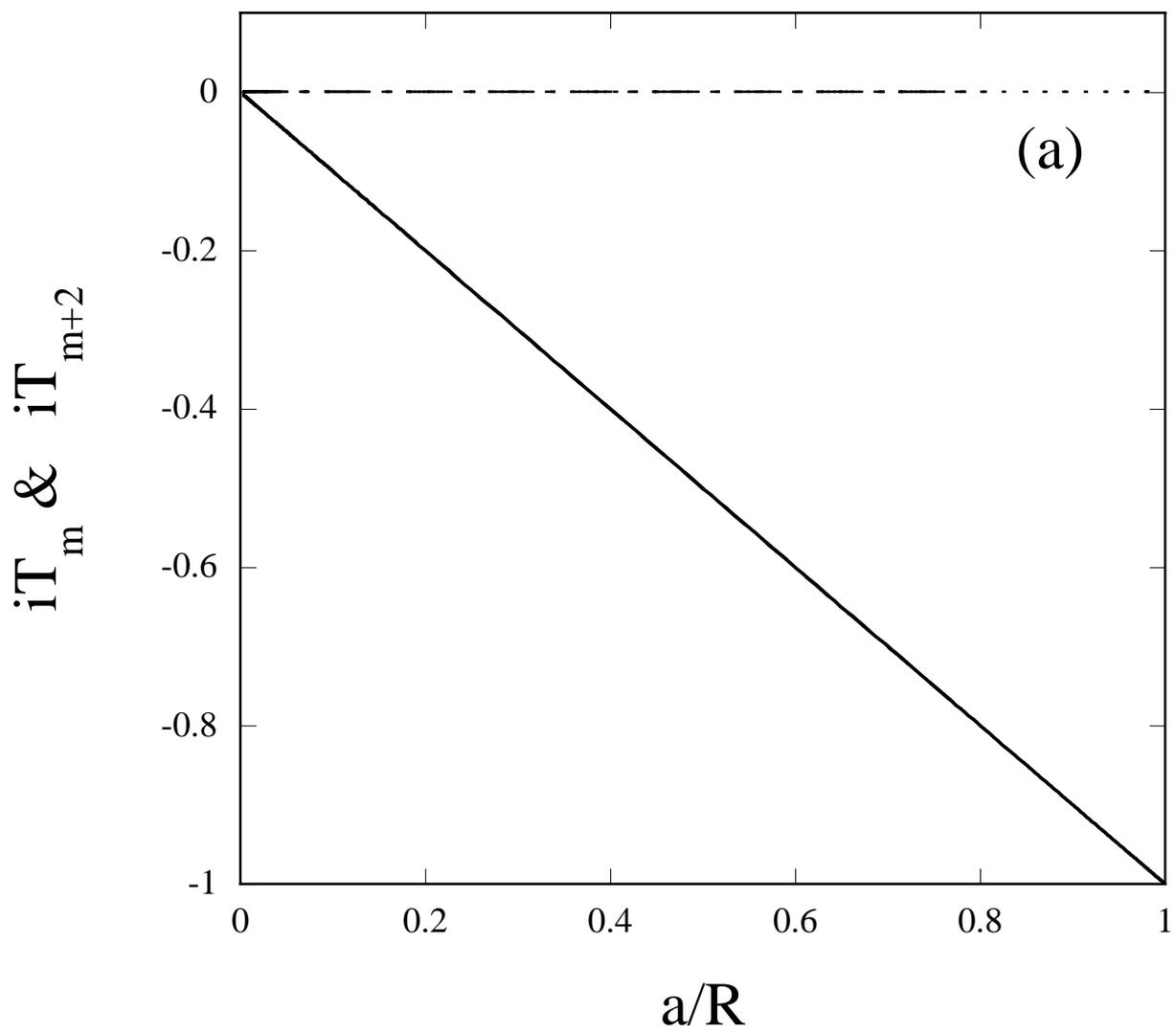
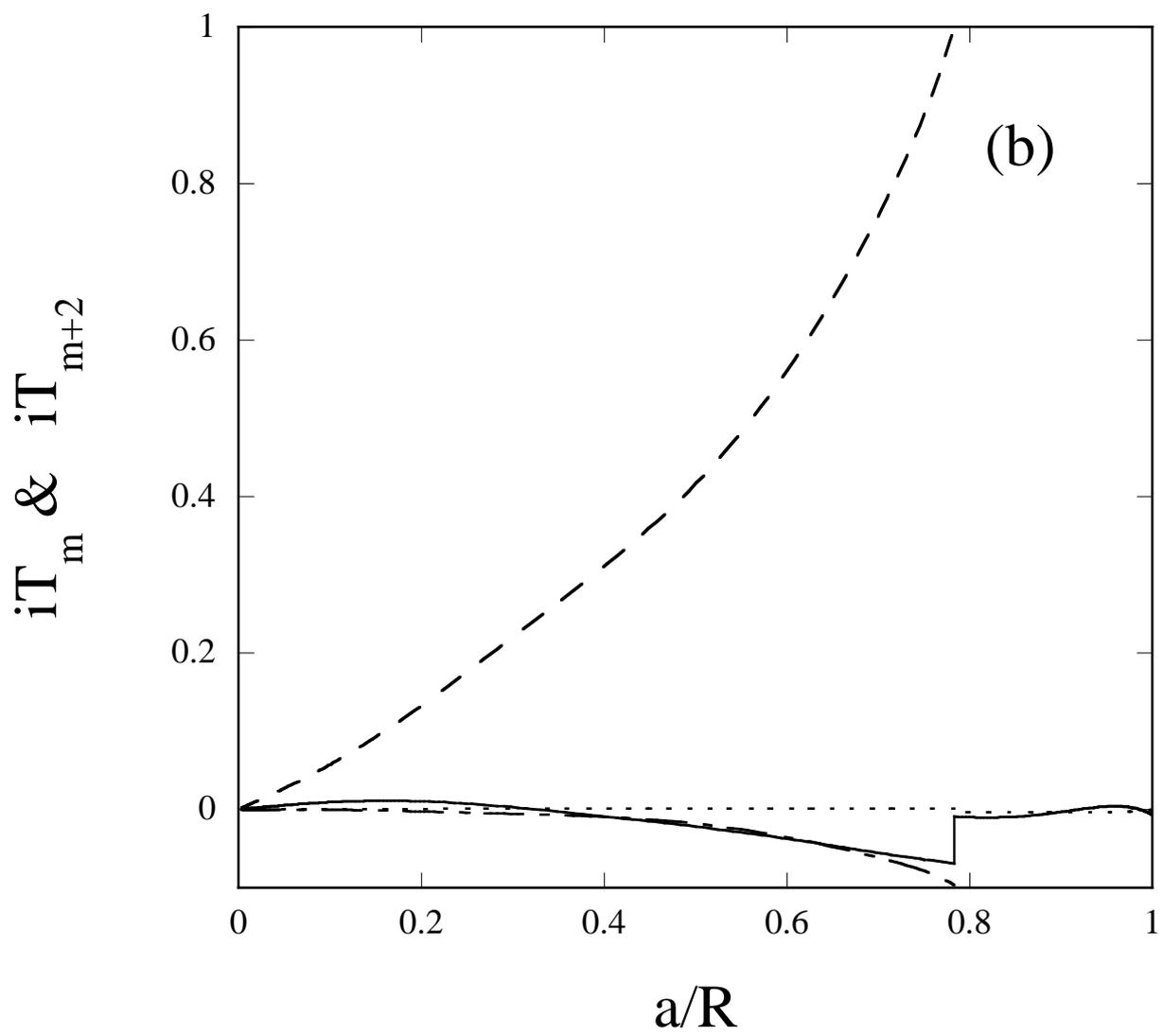
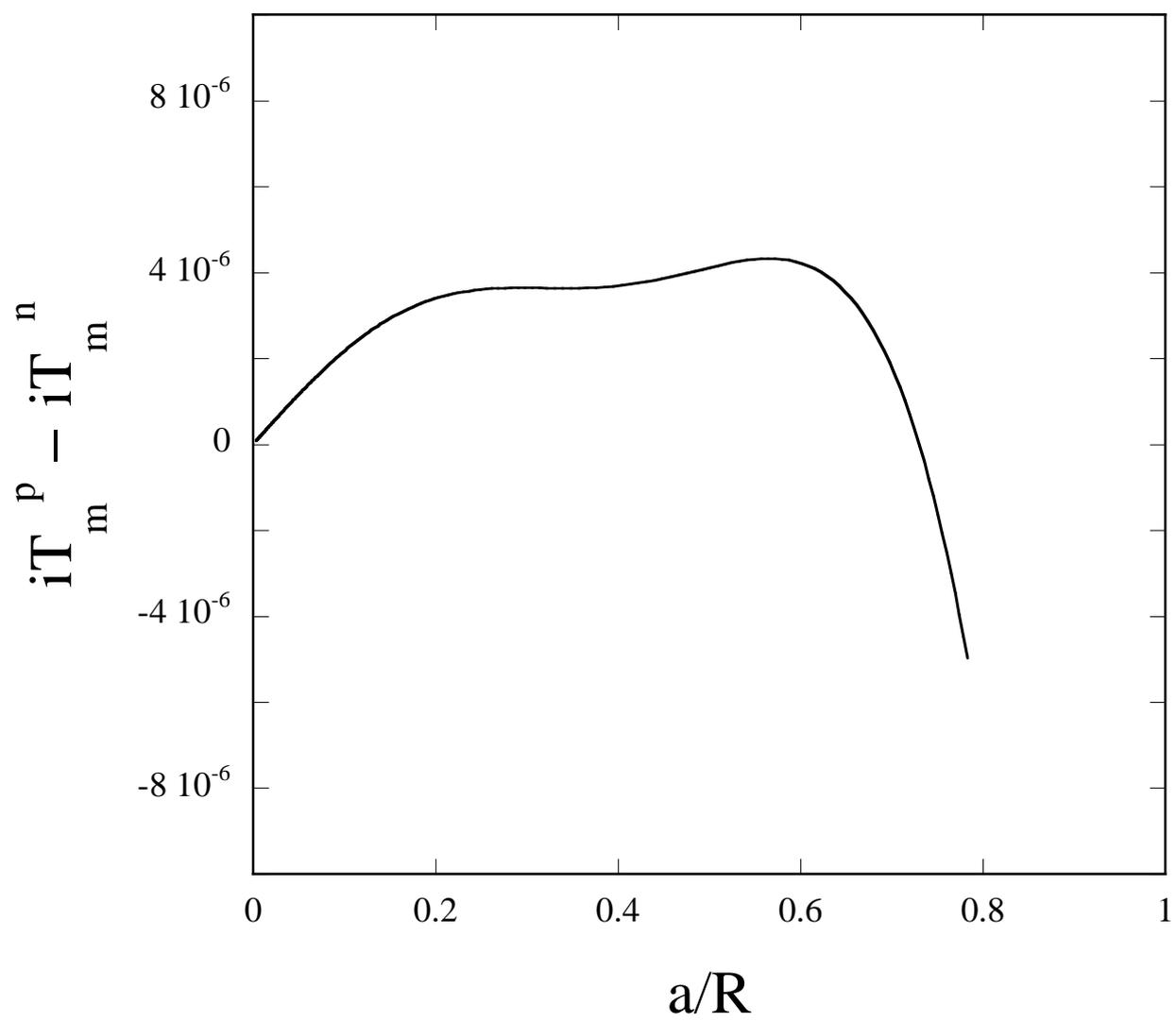
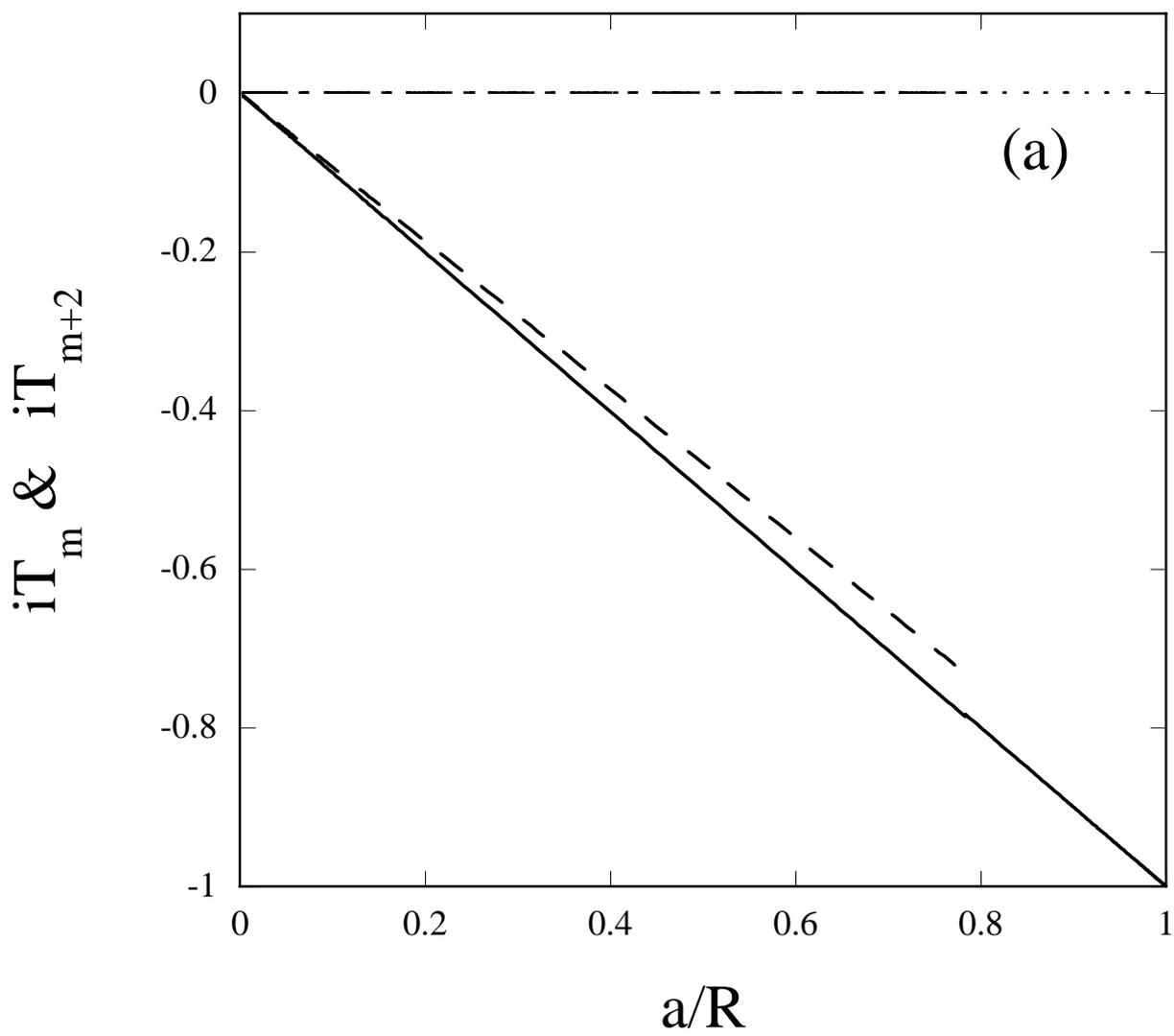
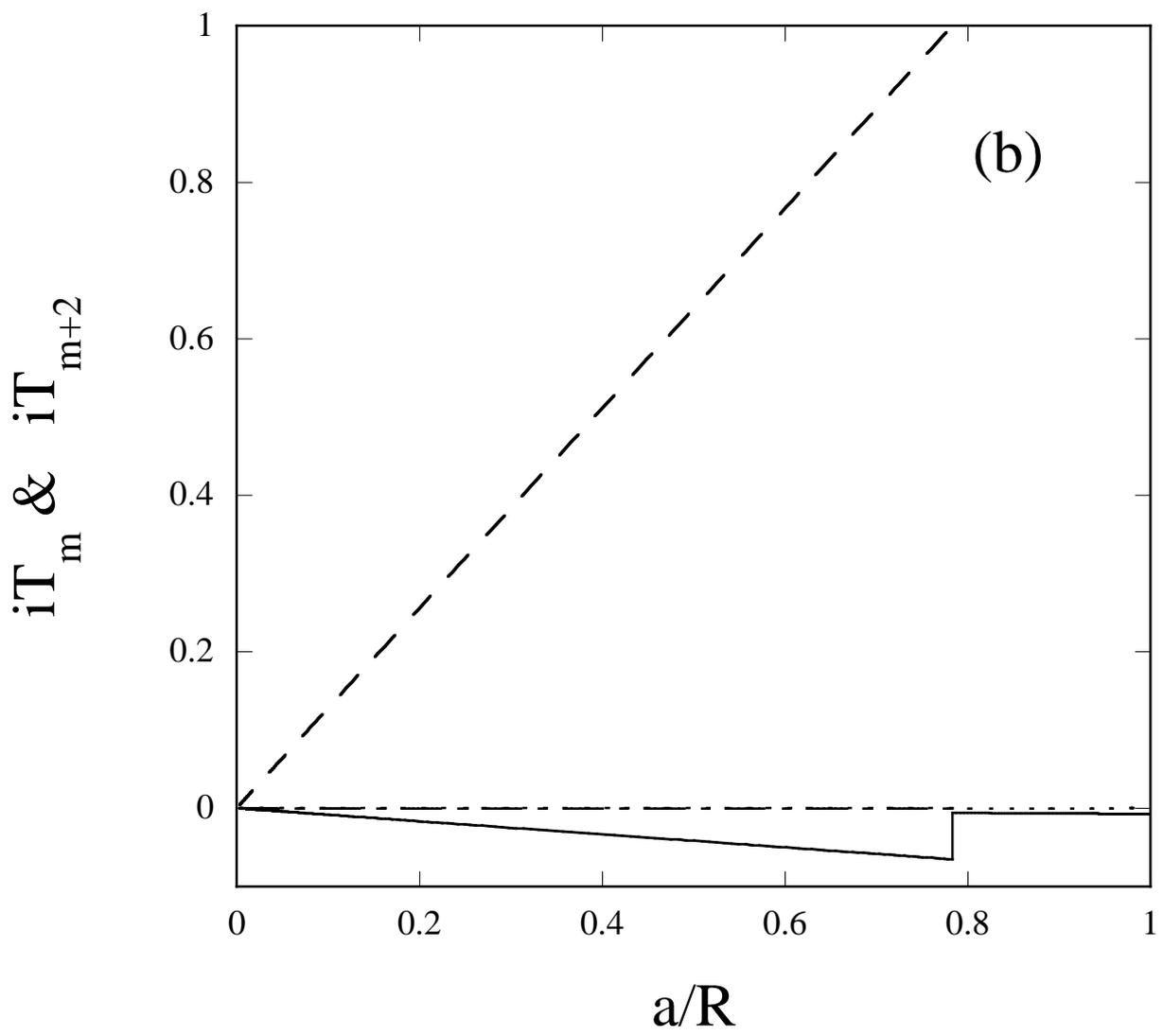
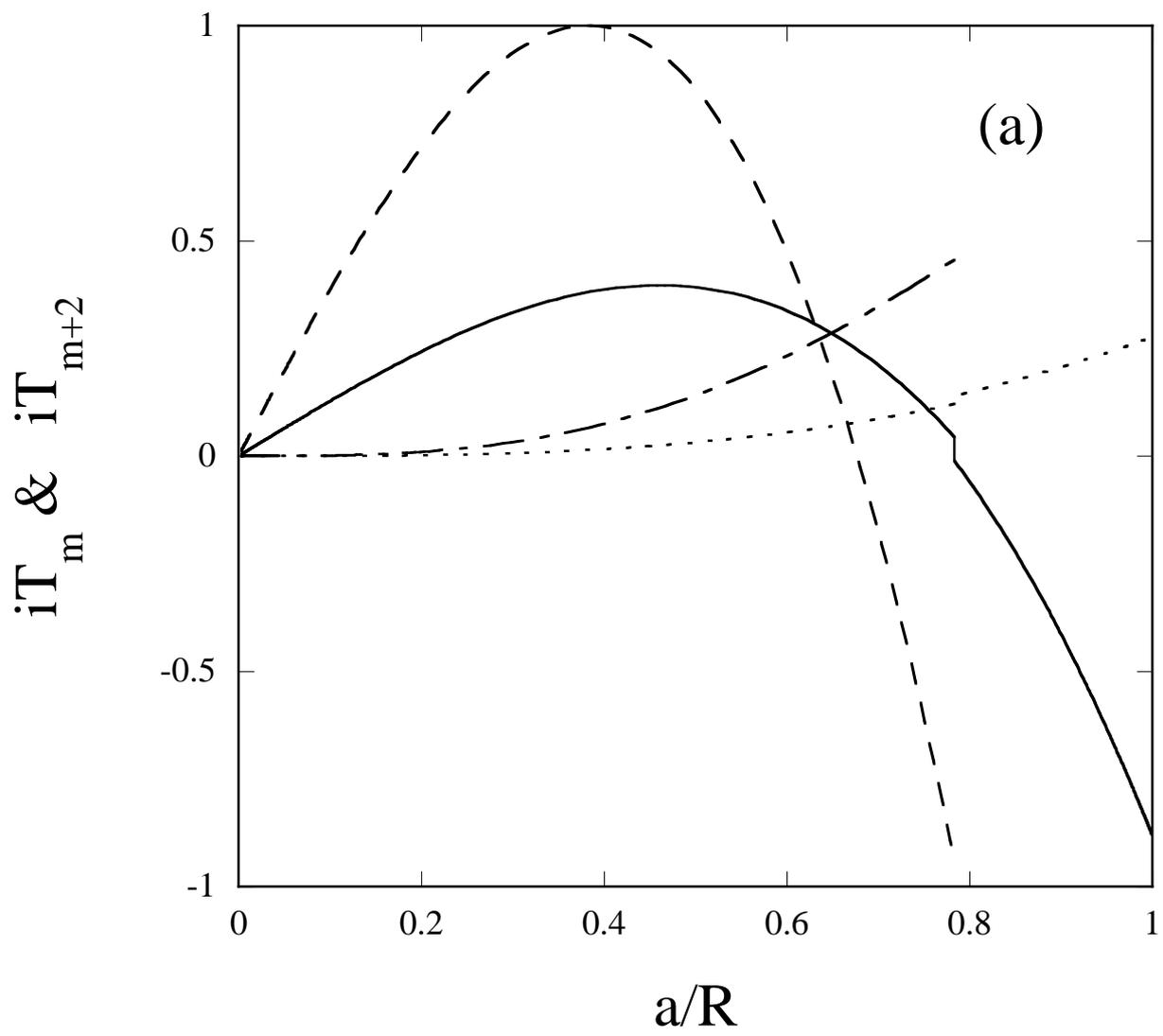
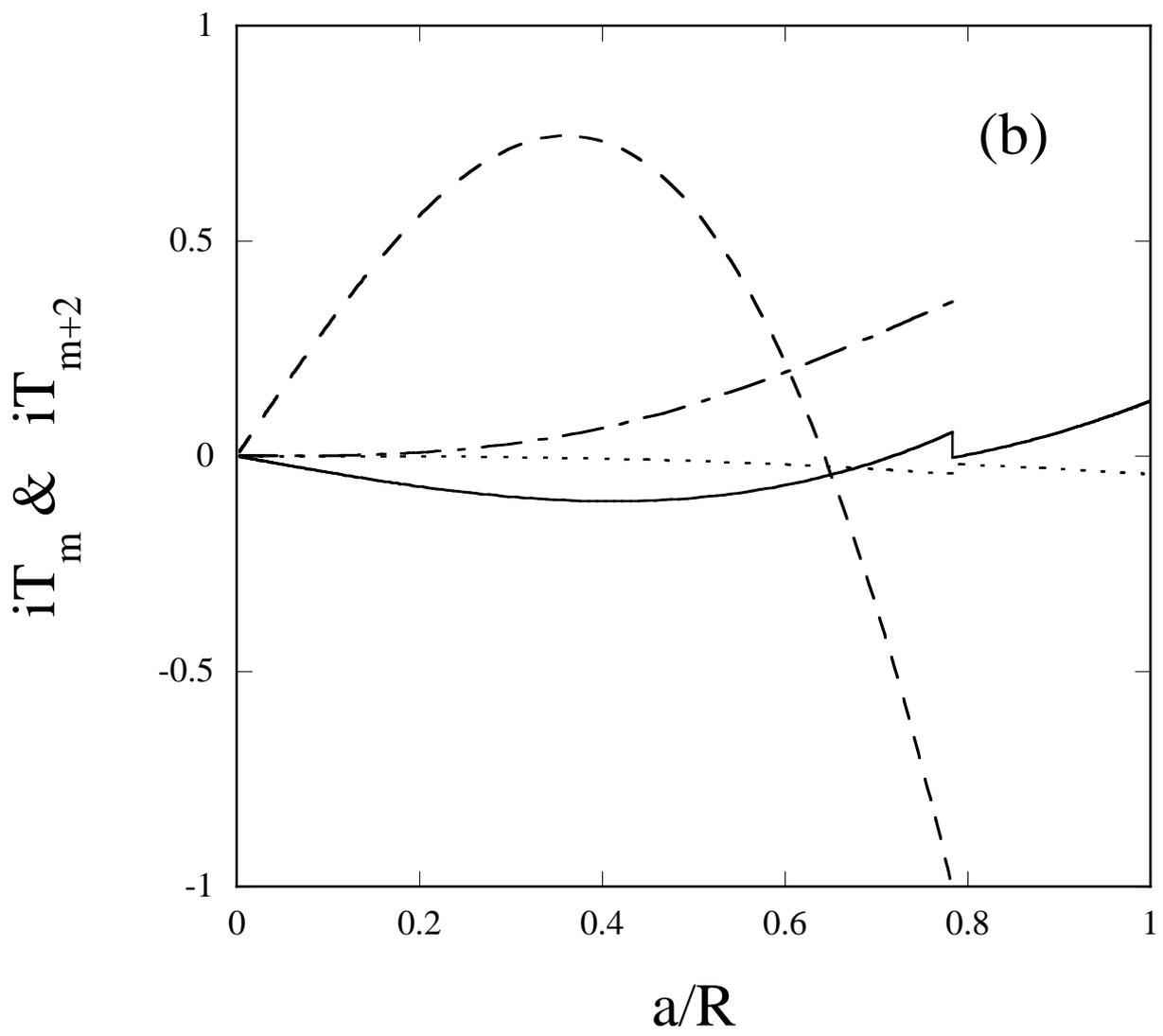
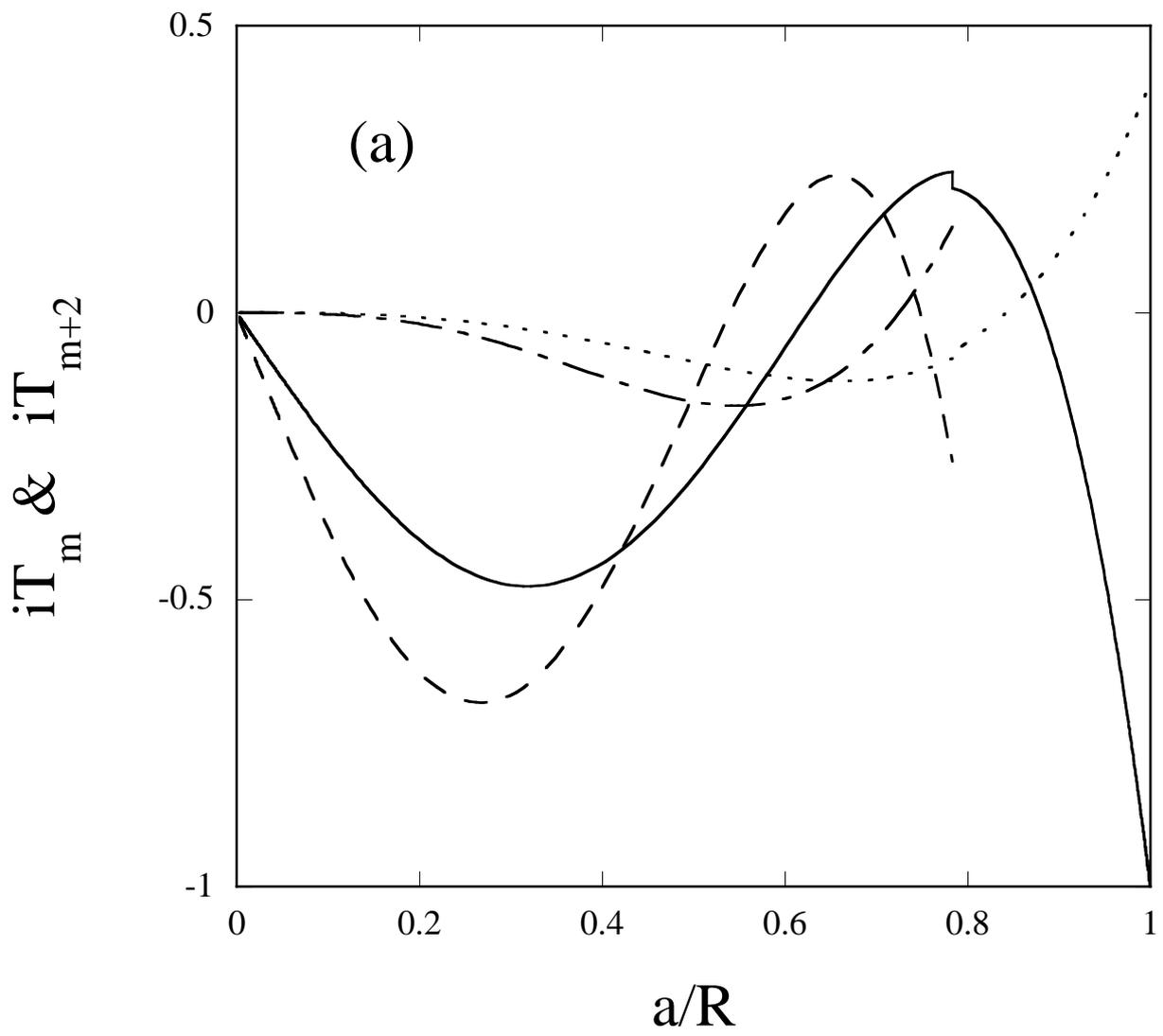
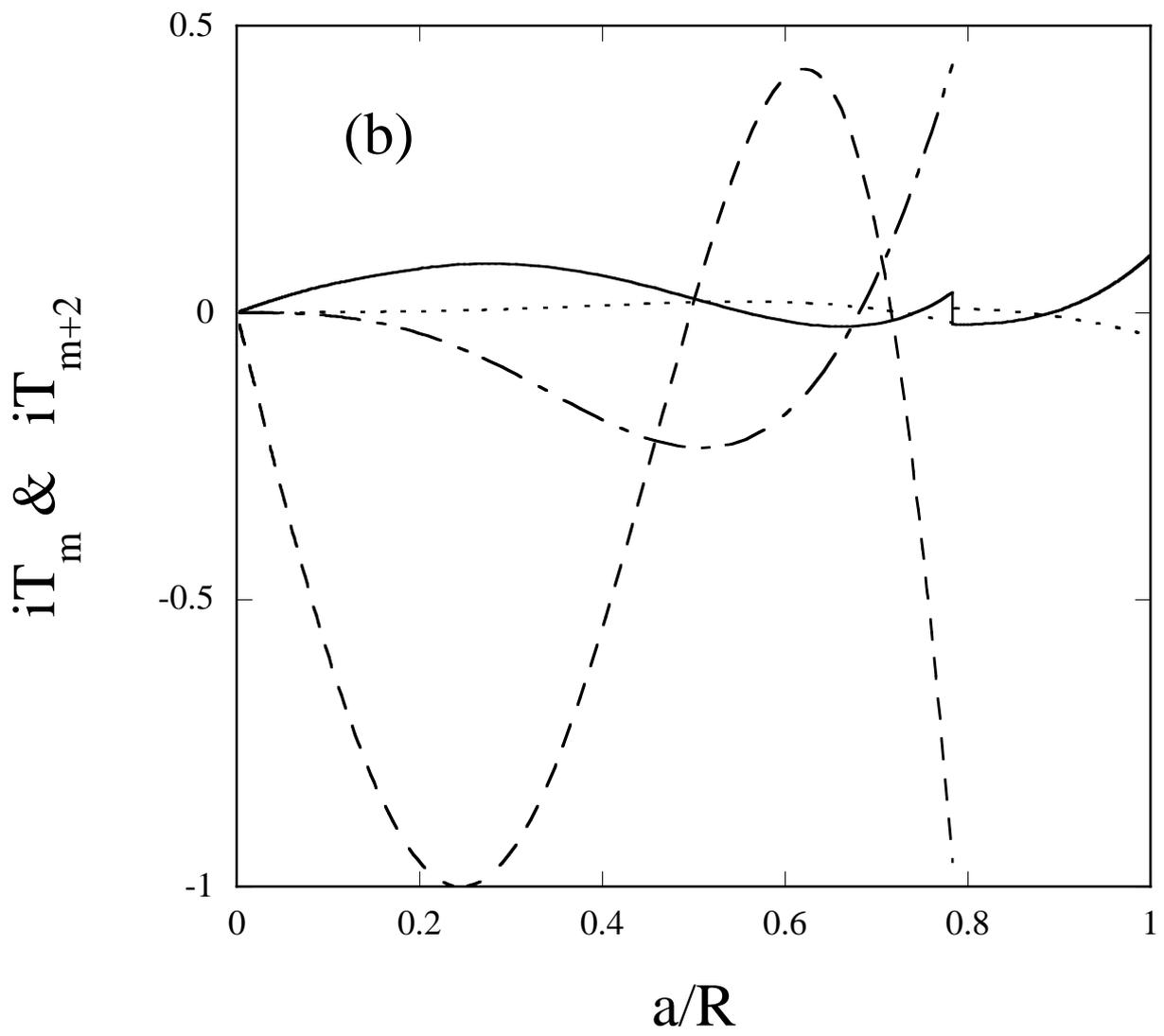
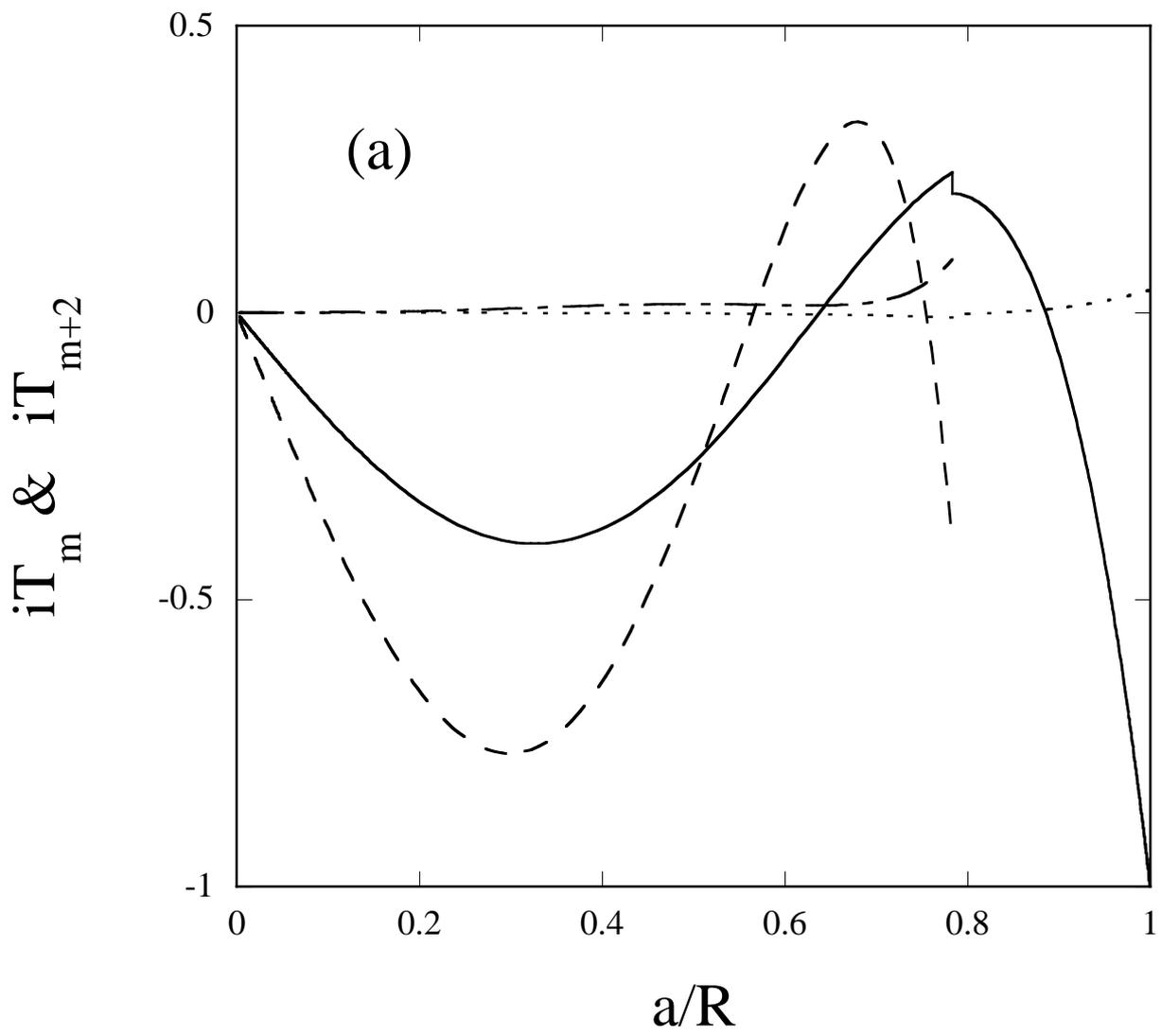
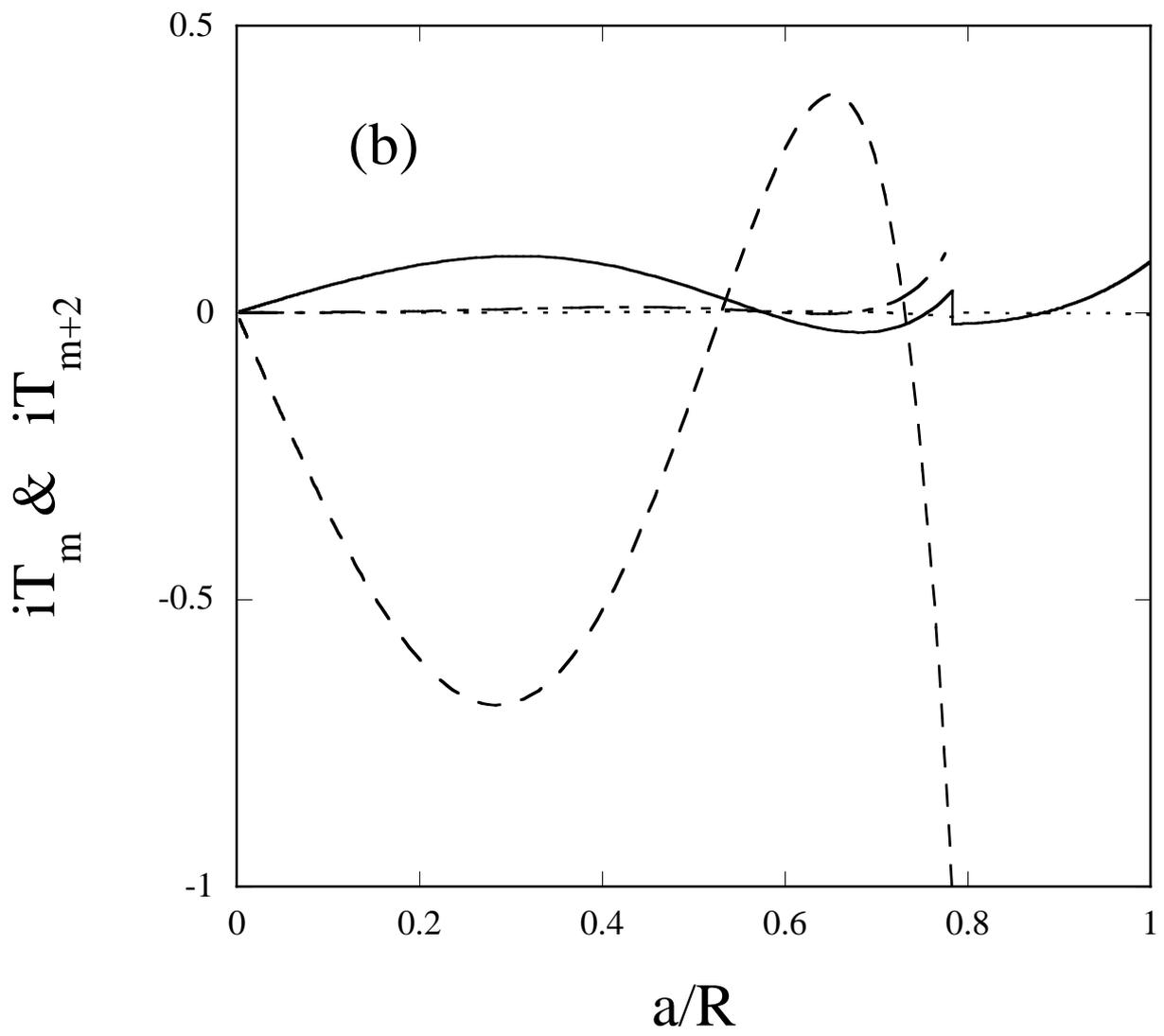
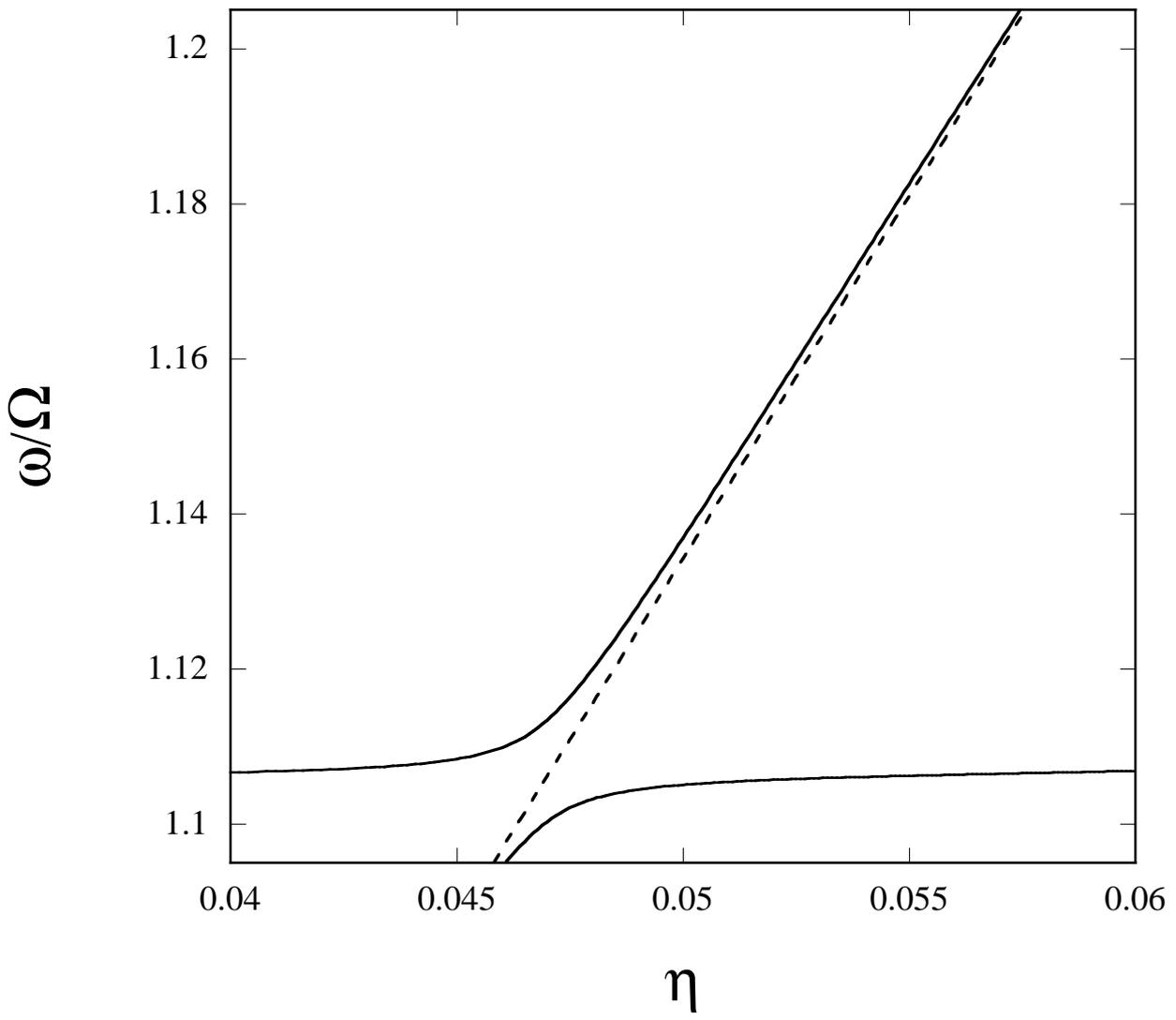
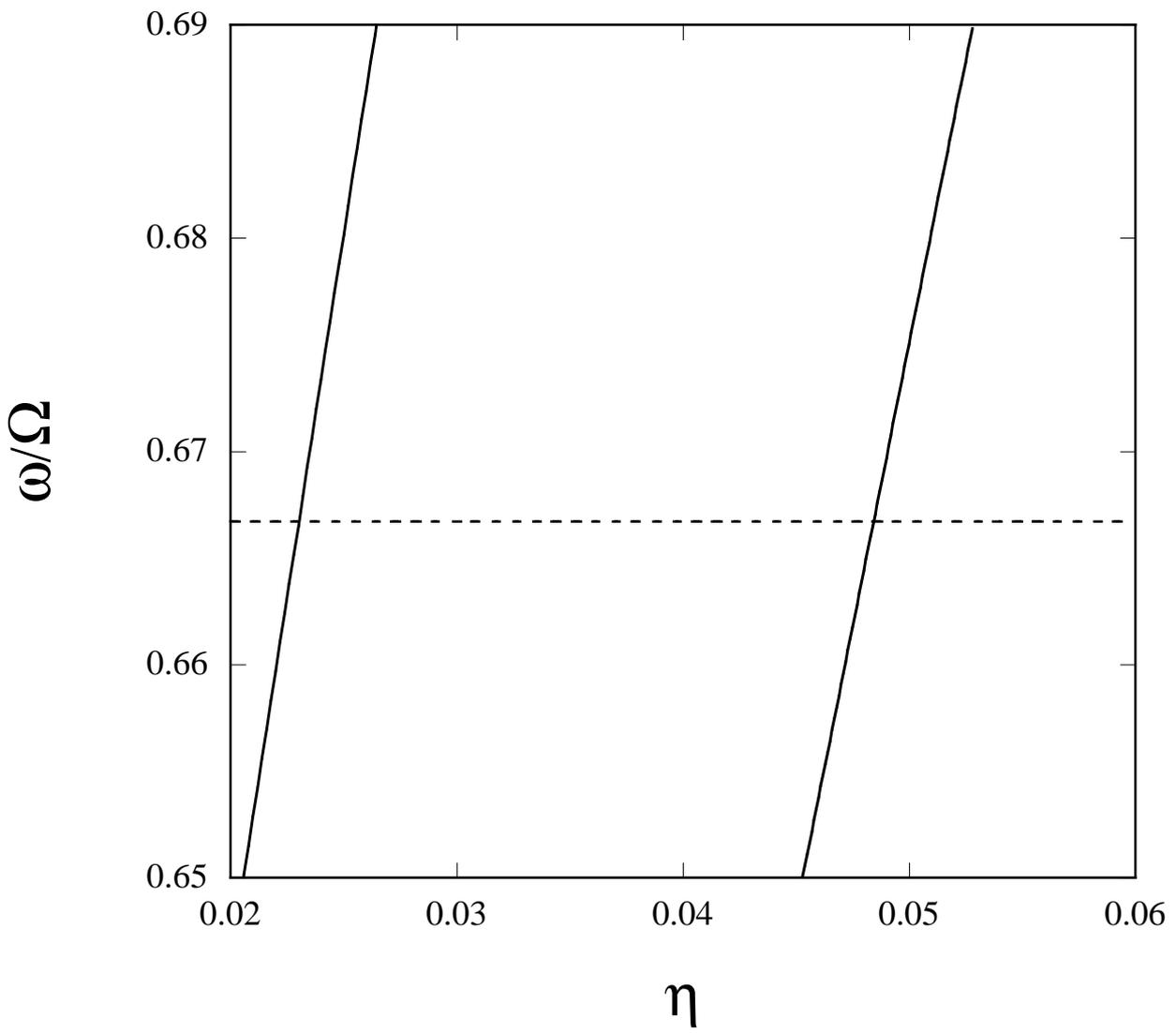
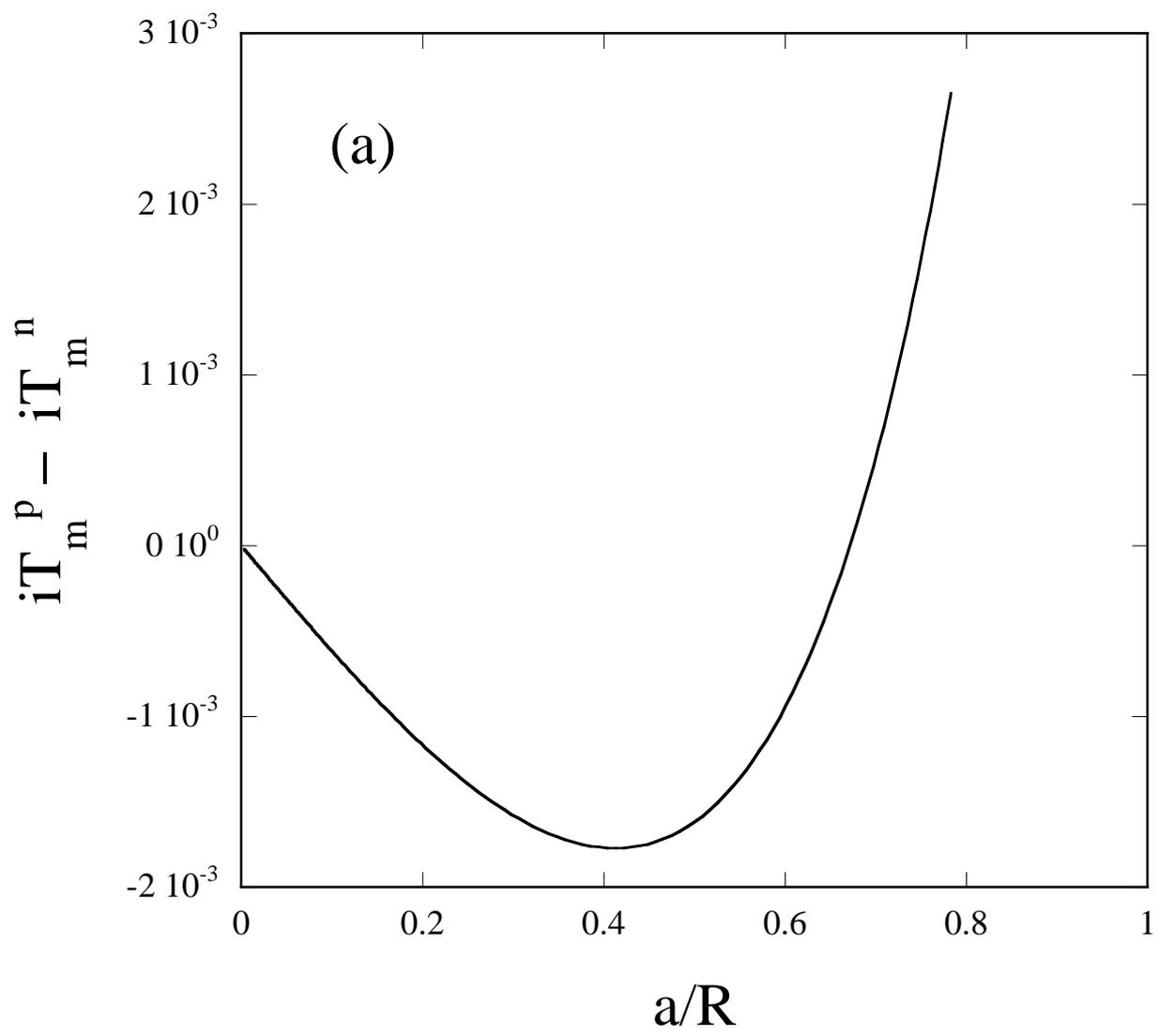
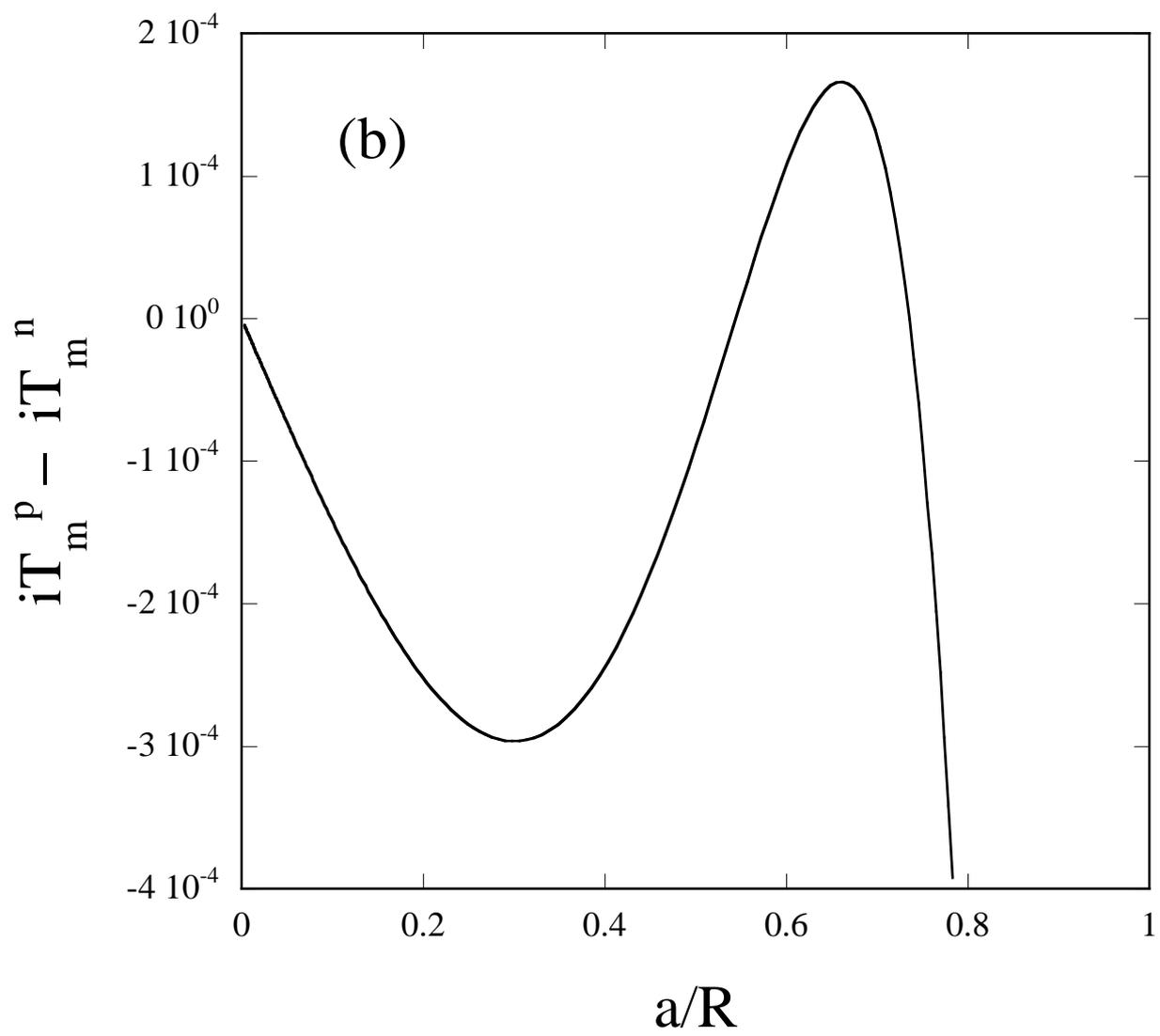
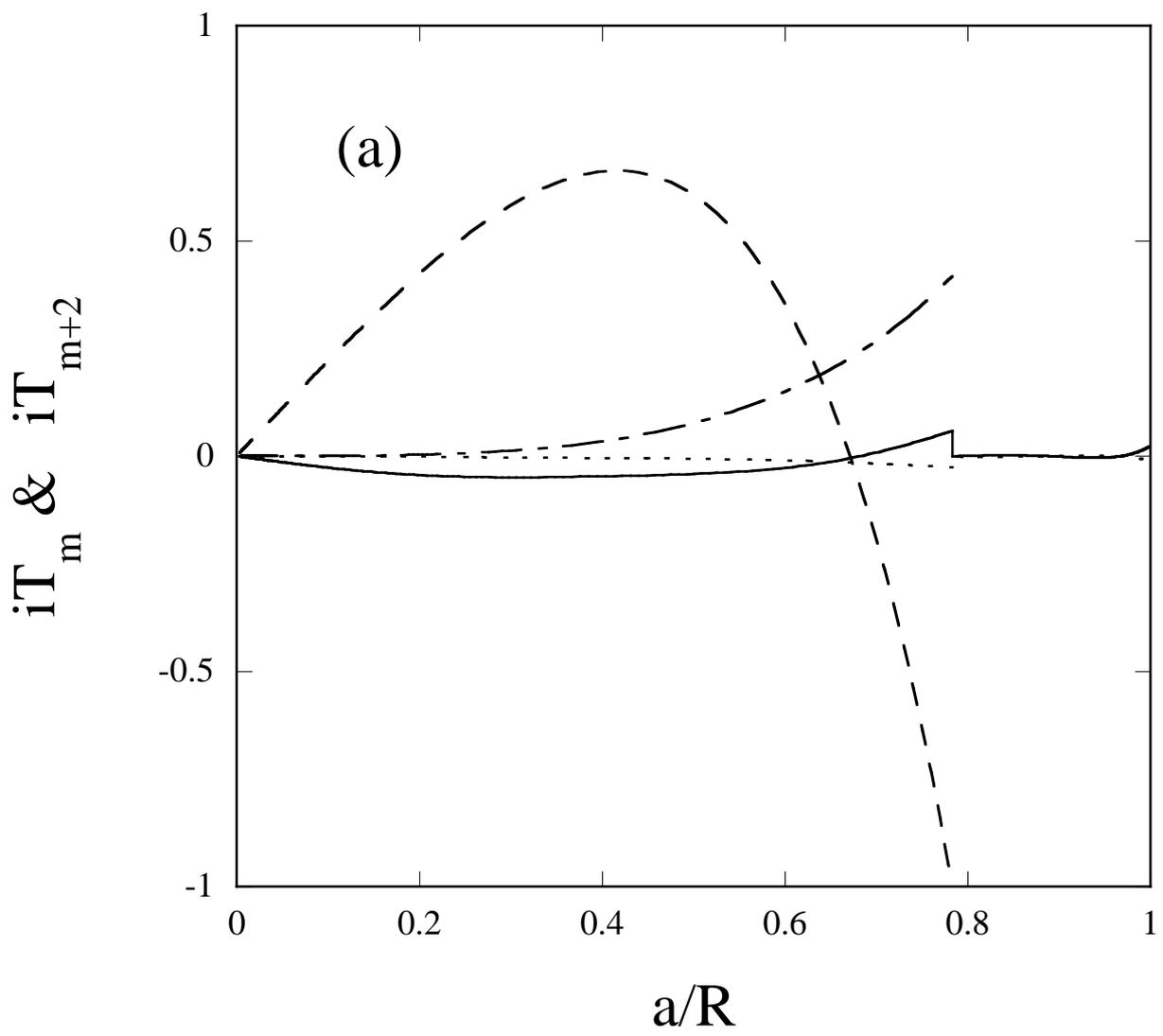
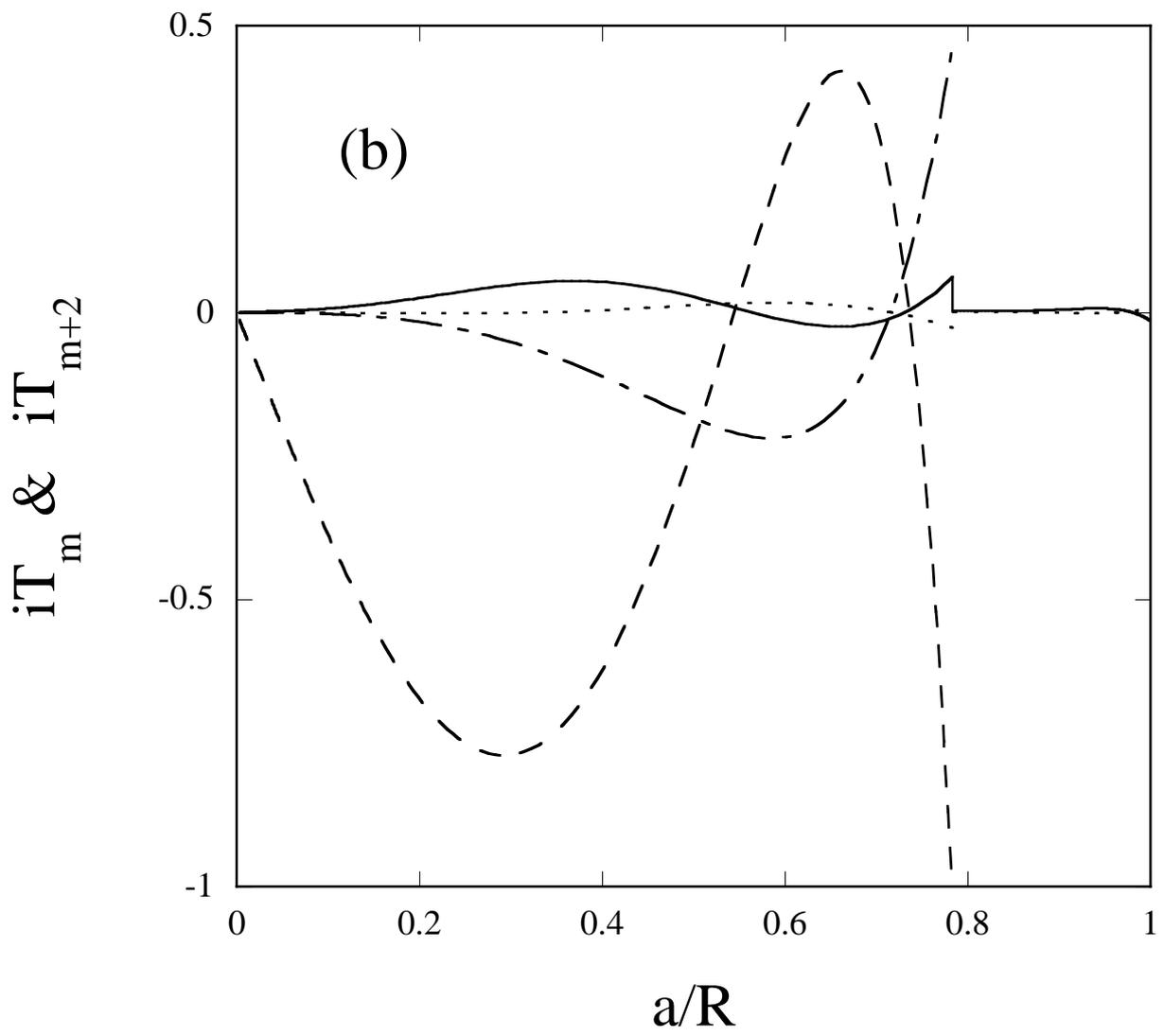
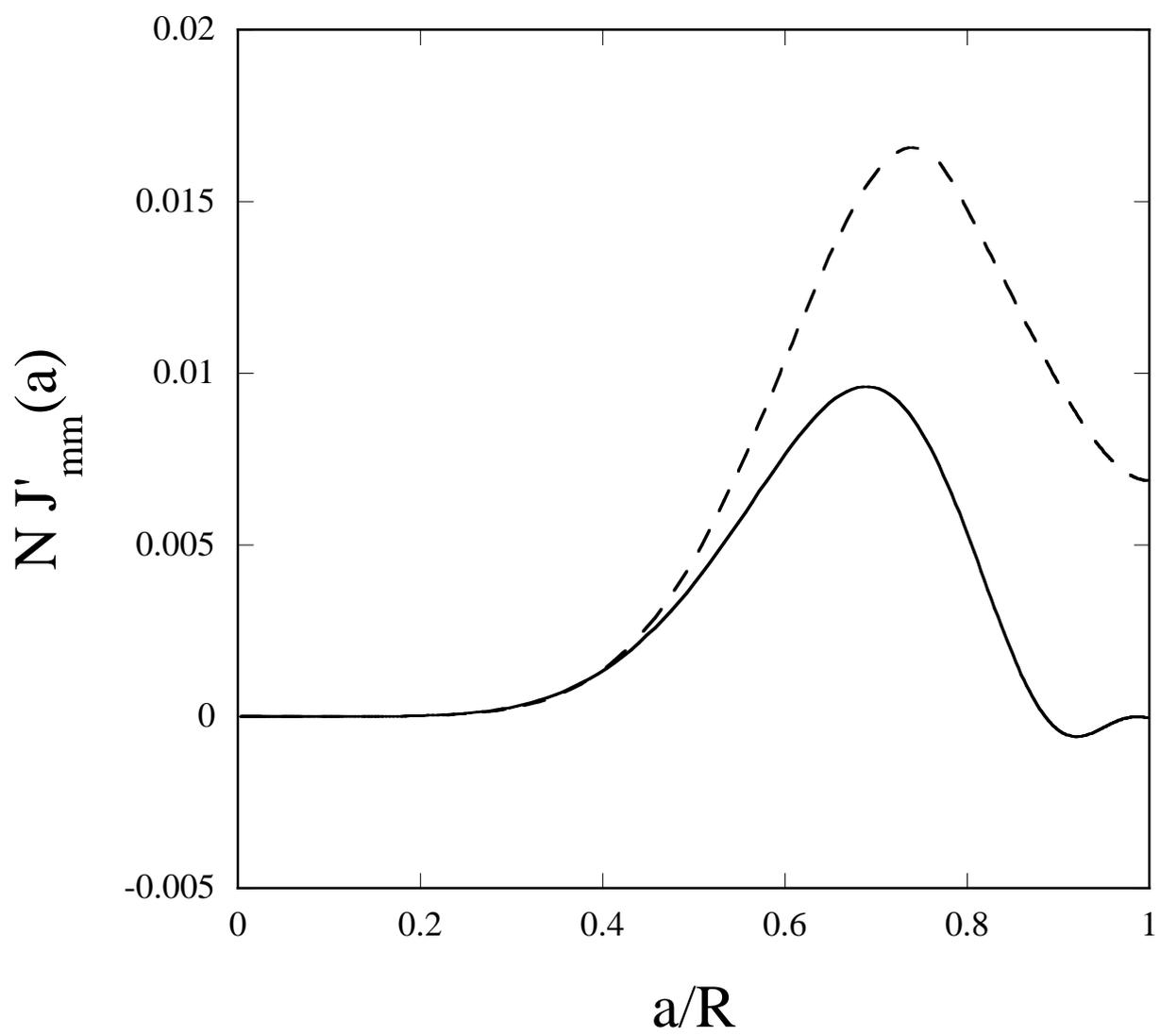
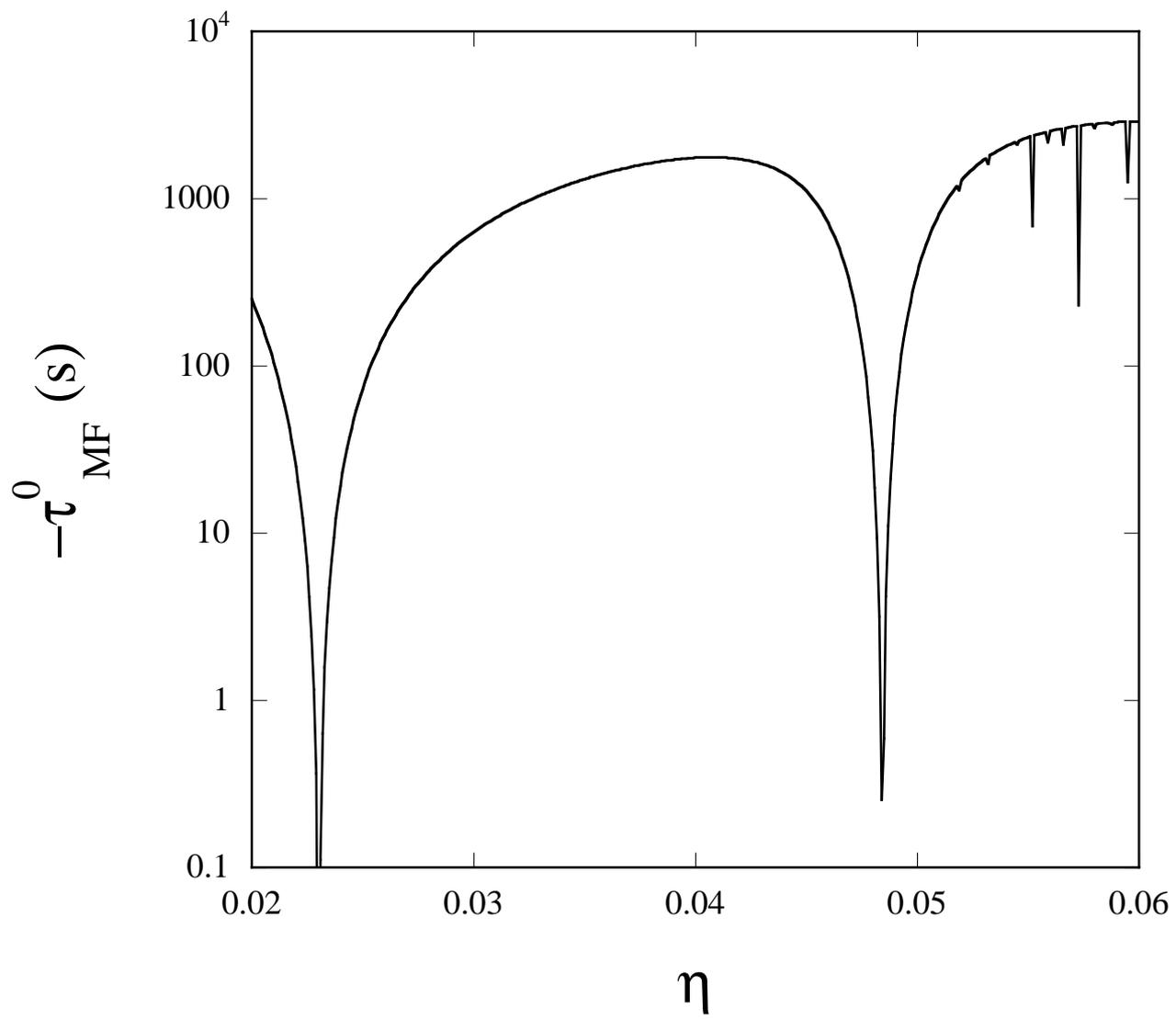
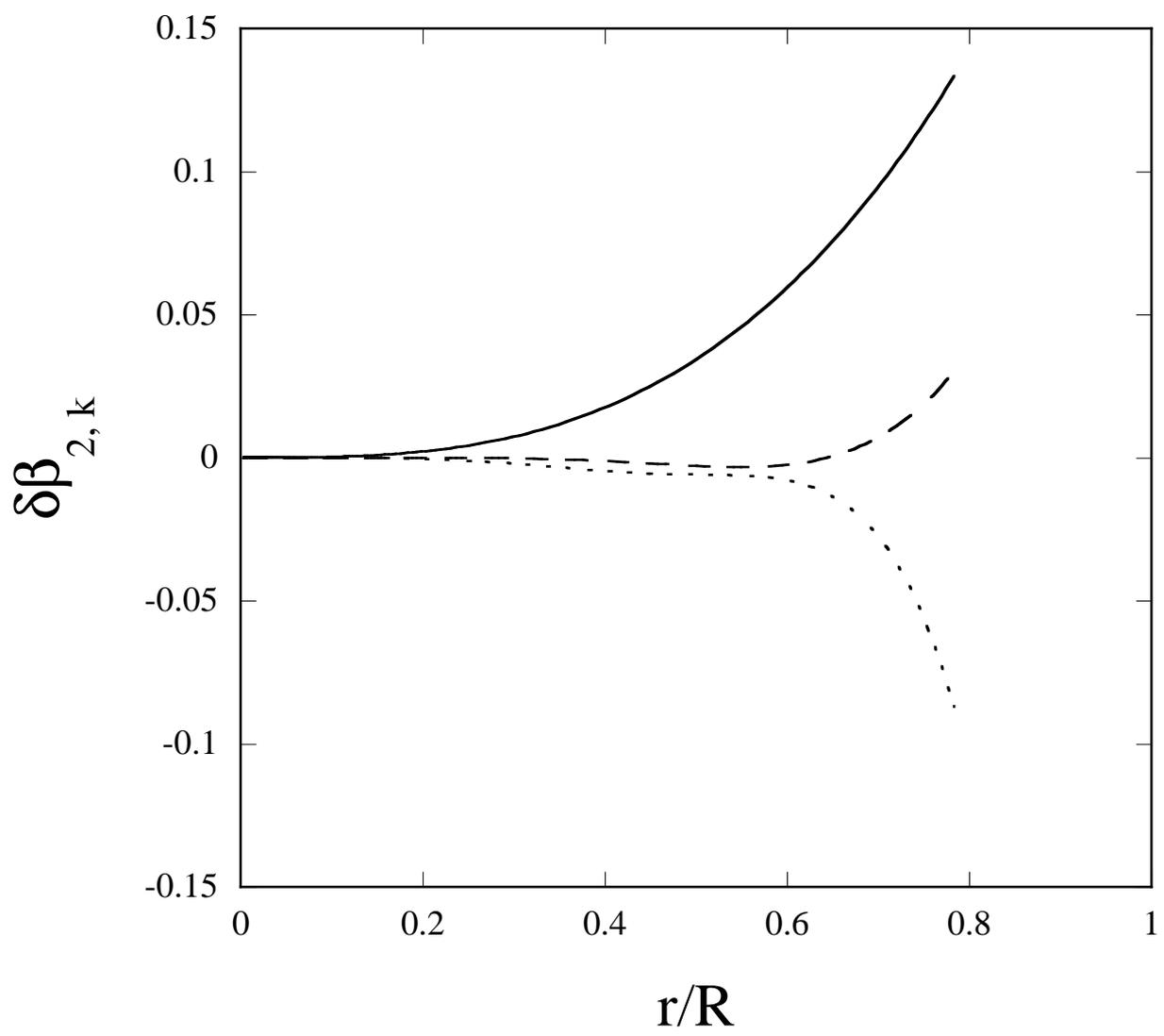
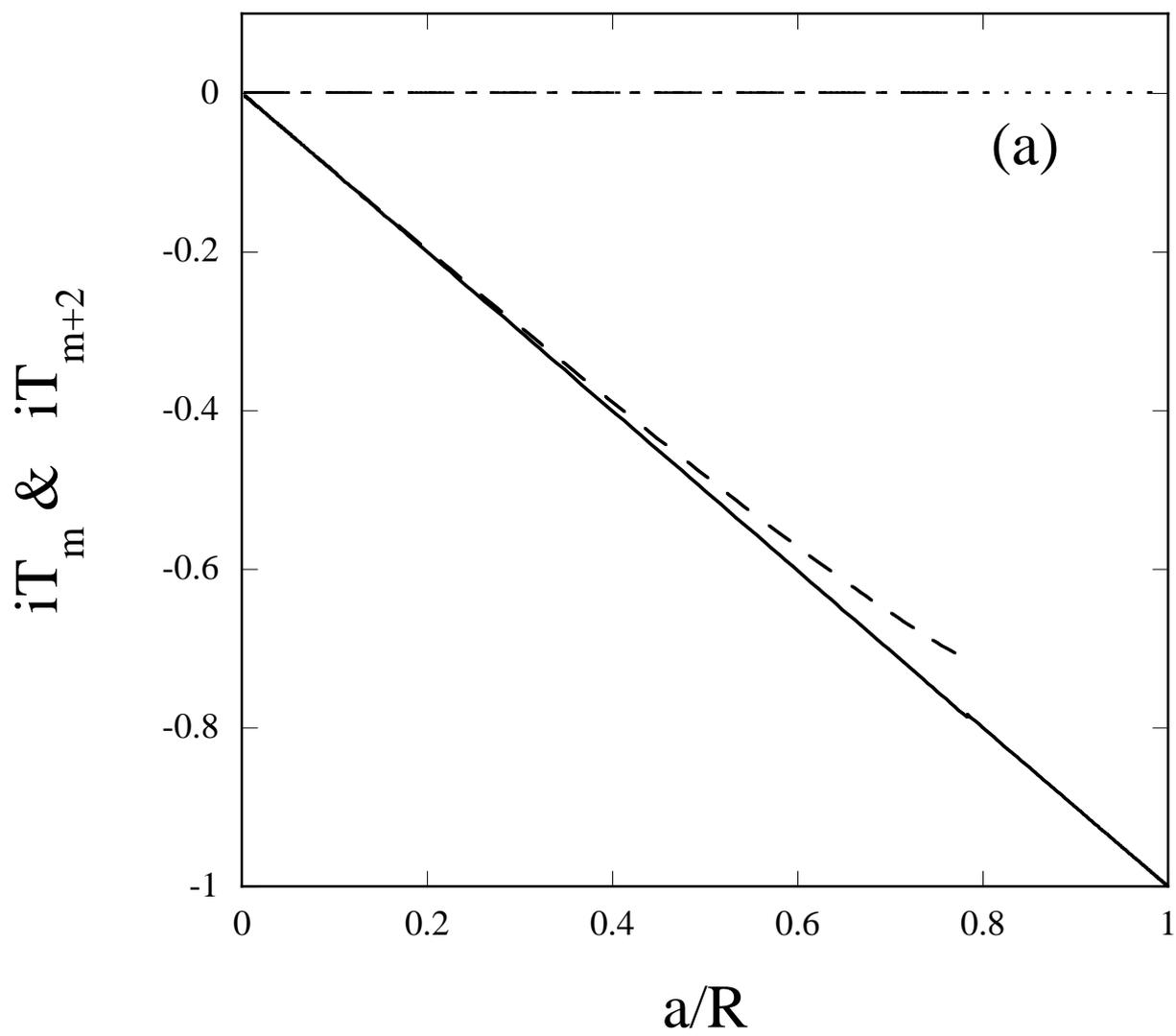
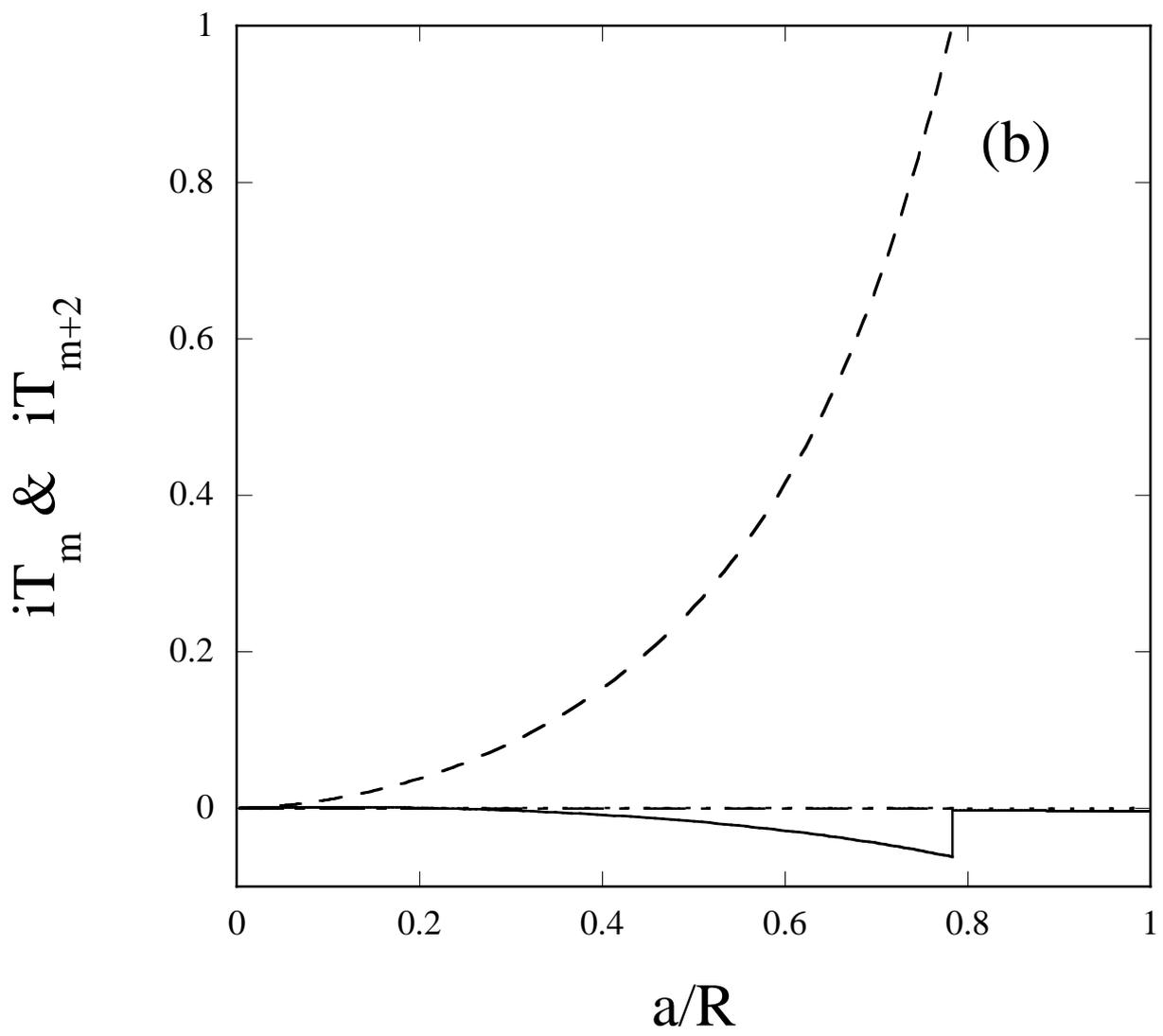
\end{document}